\documentclass{eptcs}
\usepackage{isabelle,isabellesym,amsmath,amssymb,wrapfig}
\usepackage{environ} 
\usepackage{graphicx}
\usepackage{tikz}
\usetikzlibrary{positioning}
\usetikzlibrary{arrows}
\usetikzlibrary{automata,positioning}

\newsavebox{\rofl}
\newlength\foo

\NewEnviron{isbfig}[1]{
\hfill
\begin{minipage}[c][#1]{0.25\textwidth}
    \raggedleft
    \BODY
\end{minipage}
\vskip-#1
}

\isabellestyle{it}

\begin{document}

\title{Systematic Verification of the Modal Logic Cube \\ in Isabelle/HOL\thanks{This work has been supported by the German Research Foundation DFG under grants BE2501/9-2 \&
    BE2501/11-1.}}
\author{Christoph Benzm\"uller \qquad \qquad Maximilian Claus 
\institute{Dep. of Mathematics and Computer Science, Freie
  Universit\"at Berlin, Germany}
\email{c.benzmueller|m.claus@fu-berlin.de}
\and
Nik Sultana
\institute{Computer Lab, Cambridge University, UK}
\email{nik.sultana@cl.cam.ac.uk}
}
\def\titlerunning{Systematic Verification of the Modal Logic Cube in Isabelle/HOL}
\def\authorrunning{Christoph Benzm\"uller, Maximilian Claus and Nik Sultana}

\maketitle


\parindent 0pt\parskip 0.5ex

%
\begin{isabellebody}%
\def\isabellecontext{ModalCube}%
\isadelimtheory
\endisadelimtheory
\isatagtheory
\endisatagtheory
{\isafoldtheory}%
\isadelimtheory
\endisadelimtheory
\begin{isamarkuptext}%
\begin{abstract}
We present an automated verification of the well-known
modal logic cube in Isabelle/HOL, in which we prove the inclusion relations
between the cube's logics using automated reasoning tools.
Prior work addresses this problem but without restriction to the modal logic cube,
and using encodings in first-order logic in 
combination with first-order automated theorem provers.
In contrast, our solution is
more elegant, transparent and effective. It employs an
embedding of quantified modal logic in classical higher-order
logic. Automated reasoning tools, such as Sledgehammer with LEO-II,
Satallax and CVC4, Metis and Nitpick, are employed to achieve full automation.
Though successful, the experiments also motivate some technical improvements 
in the Isabelle/HOL tool. 

\end{abstract}%
\end{isamarkuptext}%
\isamarkuptrue%
\isamarkupsection{Introduction%
}
\isamarkuptrue%
\begin{isamarkuptext}%
We present an approach to meta-reasoning about modal logics,
and apply it to verify the relative strengths of logics in the well-known \emph{modal logic cube},
which is illustrated in Figure 1. In particular, proofs are given for the
equivalences of different axiomatizations and the inclusion relations shown in the cube.

\begin{figure}[tp]
\centering

\includegraphics[width=.8\textwidth]{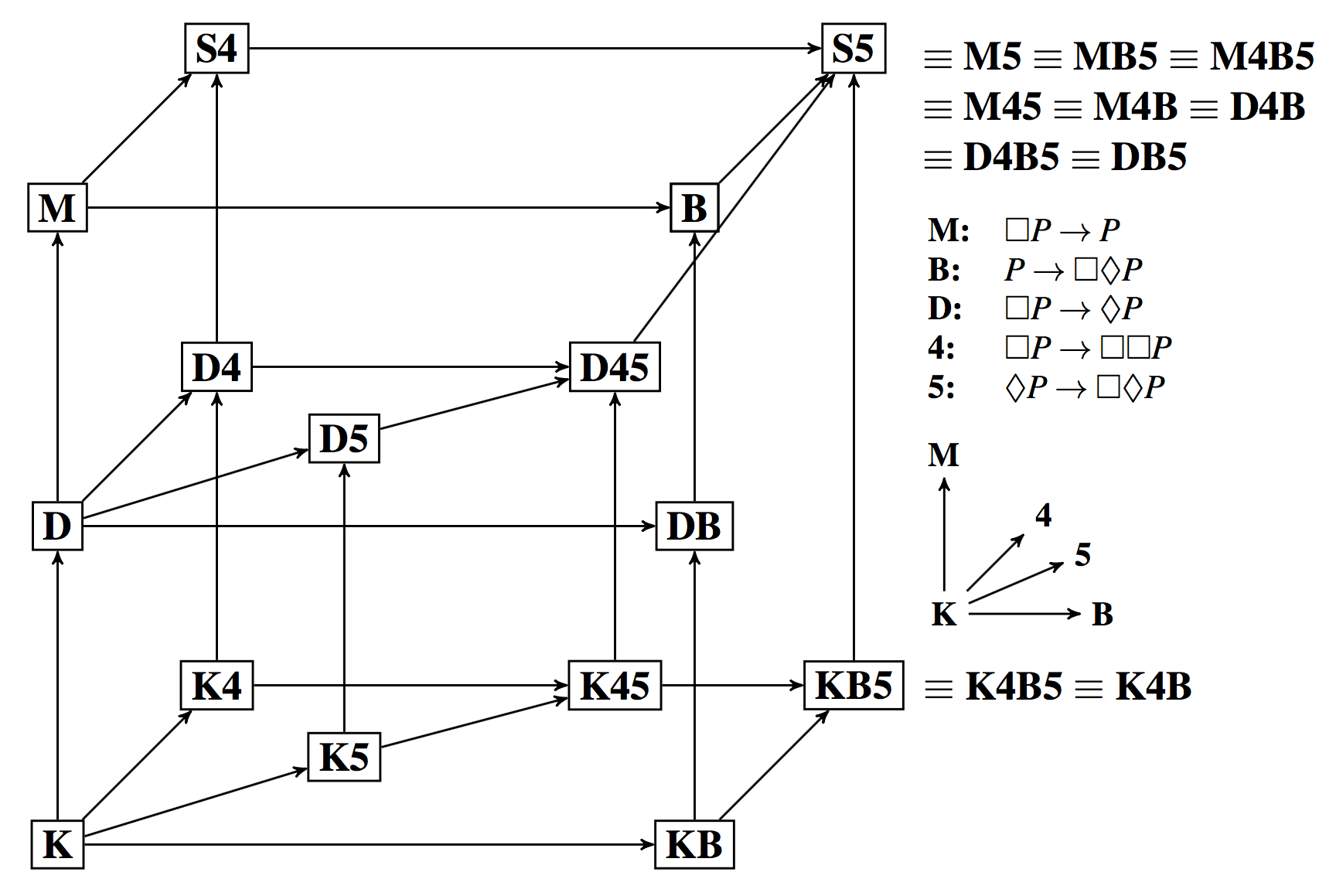}

\caption{The modal logic cube: 
reasoning in modal logics is commonly done with
respect to a certain set of basic axioms; different choices of basic
axioms give rise to different modal logics. These modal logics can be
arranged as vertices in a cube, such that the edges between them
denote inclusion relations.
}
\label{fig1}
\end{figure}

Our solution makes extensive use of the fact that all modal logics found in the cube
are sound and complete because they arise from base modal logic K by adding Sahlqvist axioms.
This is in contrast to prior work by Rabe et al.~\cite{Rabe}, who address the more general problem of determining the
relation between two arbitrary modal logics characterized by their sets of inference
rules. In their article the authors apply first-order logic encodings in combination with first-order
automated theorem provers to prove an inclusion relation employing a number of different decision strategies.
For the subproblem of only comparing logics within the cube (and therefore taking advantage of normality as additional knowledge)
our solution improves on the elegance and simplicity of the problem encodings, as well as with automation performance.
One motivation of this paper is to demonstrate the advantage of a pragmatically more expressive logic
environment (here classical higher-order logic) in comparison to a less expressive language such as
first-order logic or decidable fragments thereof.

We exploit an embedding of quantified multimodal
logic (QML) in classical higher-order logic (HOL) \cite{J23}, in which we carry out the
automated verification of the aforementioned inclusion relations. These include the logics \textbf{K}, \textbf{D},
\textbf{M} (also known as \textbf{T}), \textbf{S4}, and
\textbf{S5}. We analyze inclusion and equivalence
relations for modal logics that can be defined from normal modal logic
\textbf{K} by adding (combinations of) the axioms M, B, D, 4, and
5. In our problem encodings we exploit the well-known correspondences
between these axioms and semantic properties of accessibility
relations (i.e. Kripke models). These correspondences can themselves be elegantly formalized
and effectively automated in our approach. Formalization of the modal
axioms M, B, D, 4, and 5 requires quantification over propositional
variables. This explains why an embedding of \textit{quantified} modal
logic in HOL is needed here, and not simply an embedding of propositional
modal logic in HOL.


Our previous work (see the non-refereed, invited paper \cite{B12}) has
already demonstrated the feasibility of the approach. However, instead
of the development done there in pure TPTP THF~\cite{C25},
we here work with Isabelle/HOL~\cite{Nipkow-Paulson-Wenzel:2002} as the base
environment, and fruitfully exploit various reasoning
tools that are provided with it. This includes the
Sledgehammer-based \cite{EasyChair:128} interfaces from Isabelle/HOL to
the external higher-order theorem provers LEO-II~\cite{C26} and
Satallax~\cite{Satallax}, as well as Isabelle/HOL's own reasoner
Metis~\cite{hurd2003d}. Moreover, the higher-order model finding capabilities
of Nitpick~\cite{BlanchetteN-ITP10} are heavily used in order to formulate
and prove subsequent inclusion theorems in Isabelle/HOL.
We also encountered some problems with interacting with the proof
reconstruction available for LEO-II and Satallax in Isabelle/HOL.

This paper is a verified document in
the sense that it has been automatically generated from Isabelle/HOL
source code with the help of Isabelle's \textit{build} tool (the
entire source package is available from
\url{http://christoph-benzmueller.de/varia/pxtp2015.zip}).

The paper is structured as follows: Section~\ref{sec1} presents an
encoding of QML in HOL. This part reuses the theory provided by
Benzm\"uller and Paulson~\cite{J23}, which has recently been further
developed (to cover full higher-order QML) and applied for the verification of
G\"odel's ontological argument~\cite{GoedelGod-AFP,ECAI}. Section~\ref{sec2}
first establishes the well-known correspondence between properties of models
and base axioms, and then investigates the equivalence of different axiomatizations.
Subsequently, all inclusion relations as depicted in the modal logic cube are shown to be proper. Finally, 
the minimal number of possible worlds that is required to obtain proper inclusions in each case 
is determined and verified. Section~\ref{sec:eval} presents a short evaluation and discussion of the 
conducted experiments, and Section~\ref{sec:conc} concludes the paper.%
\end{isamarkuptext}%
\isamarkuptrue%
\isamarkupsection{An Embedding of Quantified Multimodal Logics in HOL \label{sec1}%
}
\isamarkuptrue%
\begin{isamarkuptext}%
In contrast to the monomodal case, in quantified multimodal logics both modalities \isa{{\isasymbox}} and \isa{{\isasymdiamond}}
are para\-me\-tri\-zed, such that they refer to potentially different accessibility relations. We write
\isa{{\isasymbox}\isactrlsup R} and \isa{{\isasymdiamond}\isactrlsup R} to refer to necessity and possibility wrt.\ a relation $R$. Furthermore, in terms of quantification,
we only consider the constant-domain case: this means that all possible worlds share one common domain
of discourse. More details on the embedding of QML in HOL are given in earlier work~\cite{J23,ECAI}.%
\end{isamarkuptext}%
\isamarkuptrue%
\begin{isamarkuptext}%
QML formulas are translated as HOL terms of type \isa{i\ {\isasymRightarrow}\ bool}, where \isa{i} is the type of possible worlds.
This type is abbreviated as \isa{{\isasymsigma}}.%
\end{isamarkuptext}%
\isamarkuptrue%
\begin{isamarkuptext}%
The classical connectives $\neg, \wedge, \rightarrow$, and $\forall$
(which quantifies over individuals and over sets of individuals) and $\exists$ (over individuals) are
lifted to type $\sigma$. The lifted connectives are \isa{{\isasymnot}\isactrlsup m}, \isa{{\isasymand}\isactrlsup m}, \isa{{\isasymor}\isactrlsup m}, 
\isa{{\isasymrightarrow}\isactrlsup m}, \isa{{\isasymequiv}\isactrlsup m}, \isa{{\isasymforall}}, and \isa{{\isasymexists}} (the latter two are modeled as constant symbols). 
Other connectives can be introduced analogously. Moreover, the modal 
operators \isa{{\isasymbox}} and \isa{{\isasymdiamond}}, parametric to \isa{R},  are introduced.
Note that in symbols like \isa{{\isasymnot}\isactrlsup m}, symbol \isa{m} is simply part of the name,
whereas in \isa{{\isasymbox}\isactrlsup R} and \isa{{\isasymdiamond}\isactrlsup R}, symbol \isa{R} is a parameter to the modality.%
\end{isamarkuptext}%
\isamarkuptrue%
\isacommand{abbreviation}\isamarkupfalse%
\ mnot\ {\isacharcolon}{\isacharcolon}\ {\isachardoublequoteopen}{\isasymsigma}\ {\isasymRightarrow}\ {\isasymsigma}{\isachardoublequoteclose}\ \ \isakeyword{where}\ {\isachardoublequoteopen}{\isasymnot}\isactrlsup m\ {\isasymphi}\ {\isasymequiv}\ {\isacharparenleft}{\isasymlambda}w{\isachardot}\ {\isasymnot}\ {\isasymphi}\ w{\isacharparenright}{\isachardoublequoteclose}\ \ \ \ \isanewline
\isacommand{abbreviation}\isamarkupfalse%
\ mand\ {\isacharcolon}{\isacharcolon}\ {\isachardoublequoteopen}{\isasymsigma}\ {\isasymRightarrow}\ {\isasymsigma}\ {\isasymRightarrow}\ {\isasymsigma}{\isachardoublequoteclose}\ \ \isakeyword{where}\ {\isachardoublequoteopen}{\isasymphi}\ {\isasymand}\isactrlsup m\ {\isasympsi}\ {\isasymequiv}\ {\isacharparenleft}{\isasymlambda}w{\isachardot}\ {\isasymphi}\ w\ {\isasymand}\ {\isasympsi}\ w{\isacharparenright}{\isachardoublequoteclose}\ \ \ \isanewline
\isacommand{abbreviation}\isamarkupfalse%
\ mor\ {\isacharcolon}{\isacharcolon}\ {\isachardoublequoteopen}{\isasymsigma}\ {\isasymRightarrow}\ {\isasymsigma}\ {\isasymRightarrow}\ {\isasymsigma}{\isachardoublequoteclose}\ \ \isakeyword{where}\ {\isachardoublequoteopen}{\isasymphi}\ {\isasymor}\isactrlsup m\ {\isasympsi}\ {\isasymequiv}\ {\isacharparenleft}{\isasymlambda}w{\isachardot}\ {\isasymphi}\ w\ {\isasymor}\ {\isasympsi}\ w{\isacharparenright}{\isachardoublequoteclose}\ \ \ \isanewline
\isacommand{abbreviation}\isamarkupfalse%
\ mimplies\ {\isacharcolon}{\isacharcolon}\ {\isachardoublequoteopen}{\isasymsigma}\ {\isasymRightarrow}\ {\isasymsigma}\ {\isasymRightarrow}\ {\isasymsigma}{\isachardoublequoteclose}\ \ \isakeyword{where}\ {\isachardoublequoteopen}{\isasymphi}\ {\isasymrightarrow}\isactrlsup m\ {\isasympsi}\ {\isasymequiv}\ {\isacharparenleft}{\isasymlambda}w{\isachardot}\ {\isasymphi}\ w\ {\isasymlongrightarrow}\ {\isasympsi}\ w{\isacharparenright}{\isachardoublequoteclose}\ \ \isanewline
\isacommand{abbreviation}\isamarkupfalse%
\ mequiv{\isacharcolon}{\isacharcolon}\ {\isachardoublequoteopen}{\isasymsigma}\ {\isasymRightarrow}\ {\isasymsigma}\ {\isasymRightarrow}\ {\isasymsigma}{\isachardoublequoteclose}\ \ \isakeyword{where}\ {\isachardoublequoteopen}{\isasymphi}\ {\isasymequiv}\isactrlsup m\ {\isasympsi}\ {\isasymequiv}\ {\isacharparenleft}{\isasymlambda}w{\isachardot}\ {\isasymphi}\ w\ {\isasymlongleftrightarrow}\ {\isasympsi}\ w{\isacharparenright}{\isachardoublequoteclose}\ \ \isanewline
\isacommand{abbreviation}\isamarkupfalse%
\ mforall\ {\isacharcolon}{\isacharcolon}\ {\isachardoublequoteopen}{\isacharparenleft}{\isacharprime}a\ {\isasymRightarrow}\ {\isasymsigma}{\isacharparenright}\ {\isasymRightarrow}\ {\isasymsigma}{\isachardoublequoteclose}\ \ \isakeyword{where}\ {\isachardoublequoteopen}{\isasymforall}\ {\isasymPhi}\ {\isasymequiv}\ {\isacharparenleft}{\isasymlambda}w{\isachardot}\ {\isasymforall}x{\isachardot}\ {\isasymPhi}\ x\ w{\isacharparenright}{\isachardoublequoteclose}\ \ \ \isanewline
\isacommand{abbreviation}\isamarkupfalse%
\ mexists\ {\isacharcolon}{\isacharcolon}\ {\isachardoublequoteopen}{\isacharparenleft}{\isacharprime}a\ {\isasymRightarrow}\ {\isasymsigma}{\isacharparenright}\ {\isasymRightarrow}\ {\isasymsigma}{\isachardoublequoteclose}\ \ \isakeyword{where}\ {\isachardoublequoteopen}{\isasymexists}\ {\isasymPhi}\ {\isasymequiv}\ {\isacharparenleft}{\isasymlambda}w{\isachardot}\ {\isasymexists}x{\isachardot}\ {\isasymPhi}\ x\ w{\isacharparenright}{\isachardoublequoteclose}\isanewline
\isacommand{abbreviation}\isamarkupfalse%
\ mbox\ {\isacharcolon}{\isacharcolon}\ {\isachardoublequoteopen}{\isacharparenleft}i\ {\isasymRightarrow}\ i\ {\isasymRightarrow}\ bool{\isacharparenright}\ {\isasymRightarrow}\ {\isasymsigma}\ {\isasymRightarrow}\ {\isasymsigma}{\isachardoublequoteclose}\ \ \isakeyword{where}\ {\isachardoublequoteopen}{\isasymbox}\isactrlsup R\ {\isasymphi}\ {\isasymequiv}\ {\isacharparenleft}{\isasymlambda}w{\isachardot}\ {\isasymforall}v{\isachardot}\ {\isacharparenleft}R\ w\ v{\isacharparenright}\ {\isasymlongrightarrow}\ {\isasymphi}\ v{\isacharparenright}{\isachardoublequoteclose}\isanewline
\isacommand{abbreviation}\isamarkupfalse%
\ mdia\ {\isacharcolon}{\isacharcolon}\ {\isachardoublequoteopen}{\isacharparenleft}i\ {\isasymRightarrow}\ i\ {\isasymRightarrow}\ bool{\isacharparenright}\ {\isasymRightarrow}\ {\isasymsigma}\ {\isasymRightarrow}\ {\isasymsigma}{\isachardoublequoteclose}\ \ \isakeyword{where}\ {\isachardoublequoteopen}{\isasymdiamond}\isactrlsup R\ {\isasymphi}\ {\isasymequiv}\ {\isacharparenleft}{\isasymlambda}w{\isachardot}\ {\isasymexists}v{\isachardot}\ R\ w\ v\ {\isasymand}\ {\isasymphi}\ v{\isacharparenright}{\isachardoublequoteclose}%
\begin{isamarkuptext}%
For grounding lifted formulas, the meta-predicate \isa{{\isacharbrackleft}{\isasymcdot}{\isacharbrackright}}, read \isa{valid}, is introduced.%
\end{isamarkuptext}%
\isamarkuptrue%
\isacommand{abbreviation}\isamarkupfalse%
\ valid\ {\isacharcolon}{\isacharcolon}\ {\isachardoublequoteopen}{\isasymsigma}\ {\isasymRightarrow}\ bool{\isachardoublequoteclose}\ \ \isakeyword{where}\ {\isachardoublequoteopen}{\isacharbrackleft}p{\isacharbrackright}\ {\isasymequiv}\ {\isasymforall}w{\isachardot}\ p\ w{\isachardoublequoteclose}%
\isamarkupsection{Reasoning about Modal Logics \label{sec2}%
}
\isamarkuptrue%
\isamarkupsubsection{Correspondence Results%
}
\isamarkuptrue%
\begin{isamarkuptext}%
Axioms of the modal cube correspond to constraints on the underlying accessibility relations.
These constraints are as follows:%
\end{isamarkuptext}%
\isamarkuptrue%
\isanewline
\isacommand{definition}\isamarkupfalse%
\ {\isachardoublequoteopen}refl\ {\isasymequiv}\ {\isasymlambda}R\ {\isacharcolon}{\isacharcolon}\ {\isacharparenleft}i\ {\isasymRightarrow}\ i\ {\isasymRightarrow}\ bool{\isacharparenright}{\isachardot}\ {\isasymforall}S{\isachardot}\ R\ S\ S{\isachardoublequoteclose}\ \ \ \ \ \ \ \ \ \ \ \ \ \ \ \ \ \ \ \ \ \ \ \ \ \ \ \ \ \ \ \ \ \ \ \ \ \ \ \ \ \ \ \ \ \ \ \ \ \ \ \ \ \ \ \ \ %
\isamarkupcmt{reflexivity%
}
\isanewline
\isacommand{definition}\isamarkupfalse%
\ {\isachardoublequoteopen}sym\ {\isasymequiv}\ {\isasymlambda}R\ {\isacharcolon}{\isacharcolon}\ {\isacharparenleft}i\ {\isasymRightarrow}\ i\ {\isasymRightarrow}\ bool{\isacharparenright}{\isachardot}\ {\isasymforall}S\ T{\isachardot}\ {\isacharparenleft}R\ S\ T\ {\isasymlongrightarrow}\ R\ T\ S{\isacharparenright}{\isachardoublequoteclose}\ \ \ \ \ \ \ \ \ \ \ \ \ \ \ \ \ \ \ \ \ \ \ \ \ \ \ \ \ \ \ \ %
\isamarkupcmt{symmetry%
}
\isanewline
\isacommand{definition}\isamarkupfalse%
\ {\isachardoublequoteopen}ser\ {\isasymequiv}\ {\isasymlambda}R\ {\isacharcolon}{\isacharcolon}\ {\isacharparenleft}i\ {\isasymRightarrow}\ i\ {\isasymRightarrow}\ bool{\isacharparenright}{\isachardot}\ {\isasymforall}S{\isachardot}\ {\isasymexists}T{\isachardot}\ R\ S\ T{\isachardoublequoteclose}\ \ \ \ \ \ \ \ \ \ \ \ \ \ \ \ \ \ \ \ \ \ \ \ \ \ \ \ \ \ \ \ \ \ \ \ \ \ \ \ \ \ \ \ \ \ \ \ \ \ %
\isamarkupcmt{seriality%
}
\isanewline
\isacommand{definition}\isamarkupfalse%
\ {\isachardoublequoteopen}trans\ {\isasymequiv}\ {\isasymlambda}R\ {\isacharcolon}{\isacharcolon}\ {\isacharparenleft}i\ {\isasymRightarrow}\ i\ {\isasymRightarrow}\ bool{\isacharparenright}{\isachardot}\ {\isasymforall}S\ T\ U{\isachardot}\ {\isacharparenleft}R\ S\ T\ {\isasymand}\ R\ T\ U\ {\isasymlongrightarrow}\ R\ S\ U{\isacharparenright}{\isachardoublequoteclose}\ \ \ \ \ \ \ \ \ \ \ %
\isamarkupcmt{transitivity%
}
\isanewline
\isacommand{definition}\isamarkupfalse%
\ {\isachardoublequoteopen}eucl\ {\isasymequiv}\ {\isasymlambda}R\ {\isacharcolon}{\isacharcolon}\ {\isacharparenleft}i\ {\isasymRightarrow}\ i\ {\isasymRightarrow}\ bool{\isacharparenright}{\isachardot}\ {\isasymforall}S\ T\ U{\isachardot}\ {\isacharparenleft}R\ S\ T\ {\isasymand}\ R\ S\ U\ {\isasymlongrightarrow}\ R\ T\ U{\isacharparenright}{\isachardoublequoteclose}\ \ \ \ \ \ \ \ \ \ \ \ \ %
\isamarkupcmt{Euclidean%
}
\begin{isamarkuptext}%
The corresponding axioms are defined next; note that they are parametric over accessibility 
relation $R$:%
\end{isamarkuptext}%
\isamarkuptrue%
\isacommand{definition}\isamarkupfalse%
\ {\isachardoublequoteopen}M\ {\isasymequiv}\ {\isasymlambda}R\ {\isachardot}\ valid\ {\isacharparenleft}{\isasymforall}{\isacharparenleft}{\isasymlambda}P{\isachardot}\ {\isacharparenleft}{\isasymbox}\isactrlsup R\ P{\isacharparenright}\ {\isasymrightarrow}\isactrlsup m\ P{\isacharparenright}{\isacharparenright}{\isachardoublequoteclose}\isanewline
\isacommand{definition}\isamarkupfalse%
\ {\isachardoublequoteopen}B\ {\isasymequiv}\ {\isasymlambda}R\ {\isachardot}\ valid\ {\isacharparenleft}{\isasymforall}{\isacharparenleft}{\isasymlambda}P{\isachardot}\ P\ {\isasymrightarrow}\isactrlsup m\ {\isasymbox}\isactrlsup R{\isasymdiamond}\isactrlsup R\ P{\isacharparenright}{\isacharparenright}{\isachardoublequoteclose}\isanewline
\isacommand{definition}\isamarkupfalse%
\ {\isachardoublequoteopen}D\ {\isasymequiv}\ {\isasymlambda}R\ {\isachardot}\ valid\ {\isacharparenleft}{\isasymforall}{\isacharparenleft}{\isasymlambda}P{\isachardot}\ {\isacharparenleft}{\isasymbox}\isactrlsup R\ P{\isacharparenright}\ {\isasymrightarrow}\isactrlsup m\ {\isasymdiamond}\isactrlsup R\ P{\isacharparenright}{\isacharparenright}{\isachardoublequoteclose}\isanewline
\isacommand{definition}\isamarkupfalse%
\ {\isachardoublequoteopen}IV\ {\isasymequiv}\ {\isasymlambda}R\ {\isachardot}\ valid\ {\isacharparenleft}{\isasymforall}{\isacharparenleft}{\isasymlambda}P{\isachardot}\ {\isacharparenleft}{\isasymbox}\isactrlsup R\ P{\isacharparenright}\ {\isasymrightarrow}\isactrlsup m\ {\isasymbox}\isactrlsup R{\isasymbox}\isactrlsup R\ P{\isacharparenright}{\isacharparenright}{\isachardoublequoteclose}\isanewline
\isacommand{definition}\isamarkupfalse%
\ {\isachardoublequoteopen}V\ {\isasymequiv}\ {\isasymlambda}R\ {\isachardot}\ valid\ {\isacharparenleft}{\isasymforall}{\isacharparenleft}{\isasymlambda}P{\isachardot}\ {\isacharparenleft}{\isasymdiamond}\isactrlsup R\ P{\isacharparenright}\ {\isasymrightarrow}\isactrlsup m\ {\isasymbox}\isactrlsup R{\isasymdiamond}\isactrlsup R\ P{\isacharparenright}{\isacharparenright}{\isachardoublequoteclose}%
\begin{isamarkuptext}%
We will see below that \emph{correspondence theorems} (between axioms and constraints on accessibility relations)
can be elegantly expressed in HOL by exploiting the embedding used above.
These correspondence theorems link a constraint to every axiom---for instance, $M$ is linked to $\mathit{refl}$.
Subsequently, in order to make statements about the relationship of two logics in the cube, it is sufficient to only look at the model constraints of their
respective axiomatizations. Throughout the rest of this paper, all reasoning will be done on the model-theoretic
side and then interpreted on the proof-theoretic side by the means of this correspondence.%
\end{isamarkuptext}%
\isamarkuptrue%
\isamarkupsubsubsection{Axiom M corresponds to Reflexivity%
}
\isamarkuptrue%
\isacommand{theorem}\isamarkupfalse%
\ A{\isadigit{1}}{\isacharcolon}\ {\isachardoublequoteopen}{\isacharparenleft}{\isasymforall}R{\isachardot}\ {\isacharparenleft}refl\ R{\isacharparenright}\ {\isasymlongleftrightarrow}\ {\isacharparenleft}M\ R{\isacharparenright}{\isacharparenright}{\isachardoublequoteclose}%
\isadelimproof
\ %
\endisadelimproof
\isatagproof
\isacommand{by}\isamarkupfalse%
\ {\isacharparenleft}metis\ M{\isacharunderscore}def\ refl{\isacharunderscore}def{\isacharparenright}%
\endisatagproof
{\isafoldproof}%
\isadelimproof
\endisadelimproof
\isamarkupsubsubsection{Axiom B corresponds to Symmetry%
}
\isamarkuptrue%
\isacommand{lemma}\isamarkupfalse%
\ A{\isadigit{2}}{\isacharunderscore}a{\isacharcolon}\ {\isachardoublequoteopen}{\isacharparenleft}{\isasymforall}R{\isachardot}\ {\isacharparenleft}sym\ R{\isacharparenright}\ {\isasymlongrightarrow}\ {\isacharparenleft}B\ R{\isacharparenright}{\isacharparenright}{\isachardoublequoteclose}%
\isadelimproof
\ %
\endisadelimproof
\isatagproof
\isacommand{by}\isamarkupfalse%
\ {\isacharparenleft}metis\ B{\isacharunderscore}def\ sym{\isacharunderscore}def{\isacharparenright}%
\endisatagproof
{\isafoldproof}%
\isadelimproof
\endisadelimproof
\isanewline
\isacommand{lemma}\isamarkupfalse%
\ A{\isadigit{2}}{\isacharunderscore}b{\isacharcolon}\ \ {\isachardoublequoteopen}{\isacharparenleft}{\isasymforall}R{\isachardot}\ {\isacharparenleft}B\ R{\isacharparenright}\ {\isasymlongrightarrow}\ {\isacharparenleft}sym\ R{\isacharparenright}{\isacharparenright}{\isachardoublequoteclose}%
\isadelimproof
\ %
\endisadelimproof
\isatagproof
\isacommand{by}\isamarkupfalse%
\ {\isacharparenleft}simp\ add{\isacharcolon}B{\isacharunderscore}def\ sym{\isacharunderscore}def{\isacharcomma}\ force{\isacharparenright}%
\endisatagproof
{\isafoldproof}%
\isadelimproof
\endisadelimproof
\isanewline
\isacommand{theorem}\isamarkupfalse%
\ A{\isadigit{2}}{\isacharcolon}\ {\isachardoublequoteopen}{\isacharparenleft}{\isasymforall}R{\isachardot}\ {\isacharparenleft}sym\ R{\isacharparenright}\ {\isasymlongleftrightarrow}\ {\isacharparenleft}B\ R{\isacharparenright}{\isacharparenright}{\isachardoublequoteclose}%
\isadelimproof
\ %
\endisadelimproof
\isatagproof
\isacommand{by}\isamarkupfalse%
\ {\isacharparenleft}metis\ A{\isadigit{2}}{\isacharunderscore}a\ A{\isadigit{2}}{\isacharunderscore}b{\isacharparenright}%
\endisatagproof
{\isafoldproof}%
\isadelimproof
\endisadelimproof
\isamarkupsubsubsection{Axiom D corresponds to Seriality%
}
\isamarkuptrue%
\isacommand{theorem}\isamarkupfalse%
\ A{\isadigit{3}}{\isacharcolon}\ {\isachardoublequoteopen}{\isacharparenleft}{\isasymforall}R{\isachardot}\ {\isacharparenleft}ser\ R{\isacharparenright}\ {\isasymlongleftrightarrow}\ {\isacharparenleft}D\ R{\isacharparenright}{\isacharparenright}{\isachardoublequoteclose}%
\isadelimproof
\ %
\endisadelimproof
\isatagproof
\isacommand{by}\isamarkupfalse%
\ {\isacharparenleft}metis\ D{\isacharunderscore}def\ ser{\isacharunderscore}def{\isacharparenright}%
\endisatagproof
{\isafoldproof}%
\isadelimproof
\endisadelimproof
\isamarkupsubsubsection{Axiom 4 corresponds to Transitivity%
}
\isamarkuptrue%
\isacommand{theorem}\isamarkupfalse%
\ A{\isadigit{4}}{\isacharcolon}\ {\isachardoublequoteopen}{\isacharparenleft}{\isasymforall}R{\isachardot}\ {\isacharparenleft}trans\ R{\isacharparenright}\ {\isasymlongleftrightarrow}\ {\isacharparenleft}IV\ R{\isacharparenright}{\isacharparenright}{\isachardoublequoteclose}%
\isadelimproof
\ %
\endisadelimproof
\isatagproof
\isacommand{by}\isamarkupfalse%
\ {\isacharparenleft}metis\ IV{\isacharunderscore}def\ trans{\isacharunderscore}def{\isacharparenright}%
\endisatagproof
{\isafoldproof}%
\isadelimproof
\endisadelimproof
\isamarkupsubsubsection{Axiom 5 corresponds to Euclideanness%
}
\isamarkuptrue%
\isacommand{lemma}\isamarkupfalse%
\ A{\isadigit{5}}{\isacharunderscore}a{\isacharcolon}\ {\isachardoublequoteopen}{\isacharparenleft}{\isasymforall}R{\isachardot}\ {\isacharparenleft}eucl\ R{\isacharparenright}\ {\isasymlongrightarrow}\ {\isacharparenleft}V\ R{\isacharparenright}{\isacharparenright}{\isachardoublequoteclose}%
\isadelimproof
\ %
\endisadelimproof
\isatagproof
\isacommand{by}\isamarkupfalse%
\ {\isacharparenleft}metis\ V{\isacharunderscore}def\ eucl{\isacharunderscore}def{\isacharparenright}%
\endisatagproof
{\isafoldproof}%
\isadelimproof
\endisadelimproof
\isanewline
\isacommand{lemma}\isamarkupfalse%
\ A{\isadigit{5}}{\isacharunderscore}b{\isacharcolon}\ {\isachardoublequoteopen}{\isacharparenleft}{\isasymforall}R{\isachardot}\ {\isacharparenleft}V\ R{\isacharparenright}\ {\isasymlongrightarrow}\ {\isacharparenleft}eucl\ R{\isacharparenright}{\isacharparenright}{\isachardoublequoteclose}%
\isadelimproof
\ %
\endisadelimproof
\isatagproof
\isacommand{by}\isamarkupfalse%
\ {\isacharparenleft}simp\ add{\isacharcolon}V{\isacharunderscore}def\ eucl{\isacharunderscore}def{\isacharcomma}\ force{\isacharparenright}%
\endisatagproof
{\isafoldproof}%
\isadelimproof
\endisadelimproof
\isanewline
\isacommand{theorem}\isamarkupfalse%
\ A{\isadigit{5}}{\isacharcolon}\ {\isachardoublequoteopen}{\isacharparenleft}{\isasymforall}R{\isachardot}\ {\isacharparenleft}eucl\ R{\isacharparenright}\ {\isasymlongleftrightarrow}\ {\isacharparenleft}V\ R{\isacharparenright}{\isacharparenright}{\isachardoublequoteclose}%
\isadelimproof
\ %
\endisadelimproof
\isatagproof
\isacommand{by}\isamarkupfalse%
\ {\isacharparenleft}metis\ A{\isadigit{5}}{\isacharunderscore}a\ A{\isadigit{5}}{\isacharunderscore}b{\isacharparenright}%
\endisatagproof
{\isafoldproof}%
\isadelimproof
\endisadelimproof
\isamarkupsubsection{Alternative Axiomatisations of Modal Logics%
}
\isamarkuptrue%
\begin{isamarkuptext}%
Often the same logic within the cube can be obtained through different axiomatizations.
In this section we show how to prove different axiomatizations for logic \textbf{S5} resp. \textbf{KB5} to be equivalent. Using the
correspondence theorems from the previous section, the equivalences can be elegantly formulated solely using the properties
of accessibility relations. In Subsections~\ref{M5-and-MB5} and~\ref{M5-and-M4B5} we also add the corresponding 
statements using the modal logic axioms; this could analogously be done also for the other theorems and lemmata 
presented in Sections {{3.2 and 3.3}}.

The theorems below can be solved directly by Metis when it is provided the 
minimal set of necessary definitions. Sledgehammer (with the ATPs LEO-II and Satallax or with first-order provers) can also 
quickly solve these problems, in which case the manual selection of the required definitions is not necessary.%
\end{isamarkuptext}%
\isamarkuptrue%
\isamarkupsubsubsection{M5 $\Longleftrightarrow$ MB5 \label{M5-and-MB5}%
}
\isamarkuptrue%
\isacommand{theorem}\isamarkupfalse%
\ B{\isadigit{1}}{\isacharcolon}\ {\isachardoublequoteopen}{\isasymforall}R{\isachardot}{\isacharparenleft}{\isacharparenleft}refl\ R{\isacharparenright}\ {\isasymand}\ {\isacharparenleft}eucl\ R{\isacharparenright}{\isacharparenright}\ {\isasymlongleftrightarrow}\ {\isacharparenleft}{\isacharparenleft}refl\ R{\isacharparenright}\ {\isasymand}\ {\isacharparenleft}sym\ R{\isacharparenright}\ {\isasymand}\ {\isacharparenleft}eucl\ R{\isacharparenright}{\isacharparenright}{\isachardoublequoteclose}\isanewline
\isadelimproof
\ %
\endisadelimproof
\isatagproof
\isacommand{by}\isamarkupfalse%
\ {\isacharparenleft}metis\ eucl{\isacharunderscore}def\ refl{\isacharunderscore}def\ sym{\isacharunderscore}def{\isacharparenright}%
\endisatagproof
{\isafoldproof}%
\isadelimproof
\ \isanewline
\endisadelimproof
\isacommand{theorem}\isamarkupfalse%
\ B{\isadigit{1}}{\isacharunderscore}alt{\isacharcolon}\ {\isachardoublequoteopen}{\isasymforall}R{\isachardot}{\isacharparenleft}{\isacharparenleft}M\ R{\isacharparenright}\ {\isasymand}\ {\isacharparenleft}V\ R{\isacharparenright}{\isacharparenright}\ {\isasymlongleftrightarrow}\ {\isacharparenleft}{\isacharparenleft}M\ R{\isacharparenright}\ {\isasymand}\ {\isacharparenleft}B\ R{\isacharparenright}\ {\isasymand}\ {\isacharparenleft}V\ R{\isacharparenright}{\isacharparenright}{\isachardoublequoteclose}\isanewline
\isadelimproof
\ %
\endisadelimproof
\isatagproof
\isacommand{by}\isamarkupfalse%
\ {\isacharparenleft}metis\ A{\isadigit{1}}\ A{\isadigit{2}}\ A{\isadigit{5}}\ B{\isadigit{1}}{\isacharparenright}%
\endisatagproof
{\isafoldproof}%
\isadelimproof
\endisadelimproof
\isamarkupsubsubsection{M5 $\Longleftrightarrow$ M4B5 \label{M5-and-M4B5}%
}
\isamarkuptrue%
\isacommand{theorem}\isamarkupfalse%
\ B{\isadigit{2}}{\isacharcolon}\ {\isachardoublequoteopen}{\isasymforall}R{\isachardot}{\isacharparenleft}{\isacharparenleft}refl\ R{\isacharparenright}\ {\isasymand}\ {\isacharparenleft}eucl\ R{\isacharparenright}{\isacharparenright}\ {\isasymlongleftrightarrow}\ {\isacharparenleft}{\isacharparenleft}refl\ R{\isacharparenright}\ {\isasymand}\ {\isacharparenleft}trans\ R{\isacharparenright}\ {\isasymand}\ {\isacharparenleft}sym\ R{\isacharparenright}\ {\isasymand}\ {\isacharparenleft}eucl\ R{\isacharparenright}{\isacharparenright}{\isachardoublequoteclose}\isanewline
\isadelimproof
\ %
\endisadelimproof
\isatagproof
\isacommand{by}\isamarkupfalse%
\ {\isacharparenleft}metis\ eucl{\isacharunderscore}def\ refl{\isacharunderscore}def\ trans{\isacharunderscore}def\ sym{\isacharunderscore}def{\isacharparenright}%
\endisatagproof
{\isafoldproof}%
\isadelimproof
\isanewline
\endisadelimproof
\isacommand{theorem}\isamarkupfalse%
\ B{\isadigit{2}}{\isacharunderscore}alt{\isacharcolon}\ {\isachardoublequoteopen}{\isasymforall}R{\isachardot}{\isacharparenleft}{\isacharparenleft}M\ R{\isacharparenright}\ {\isasymand}\ {\isacharparenleft}V\ R{\isacharparenright}{\isacharparenright}\ {\isasymlongleftrightarrow}\ {\isacharparenleft}{\isacharparenleft}M\ R{\isacharparenright}\ {\isasymand}\ {\isacharparenleft}IV\ R{\isacharparenright}\ {\isasymand}\ {\isacharparenleft}B\ R{\isacharparenright}\ {\isasymand}\ {\isacharparenleft}V\ R{\isacharparenright}{\isacharparenright}{\isachardoublequoteclose}\isanewline
\isadelimproof
\ %
\endisadelimproof
\isatagproof
\isacommand{by}\isamarkupfalse%
\ {\isacharparenleft}metis\ A{\isadigit{1}}\ A{\isadigit{4}}\ A{\isadigit{5}}\ B{\isadigit{1}}{\isacharunderscore}alt\ B{\isadigit{2}}{\isacharparenright}%
\endisatagproof
{\isafoldproof}%
\isadelimproof
\endisadelimproof
\isamarkupsubsubsection{M5 $\Longleftrightarrow$ M45%
}
\isamarkuptrue%
\isacommand{theorem}\isamarkupfalse%
\ B{\isadigit{3}}{\isacharcolon}\ {\isachardoublequoteopen}{\isasymforall}R{\isachardot}{\isacharparenleft}{\isacharparenleft}refl\ R{\isacharparenright}\ {\isasymand}\ {\isacharparenleft}eucl\ R{\isacharparenright}{\isacharparenright}\ {\isasymlongleftrightarrow}\ {\isacharparenleft}{\isacharparenleft}refl\ R{\isacharparenright}\ {\isasymand}\ {\isacharparenleft}trans\ R{\isacharparenright}\ {\isasymand}\ {\isacharparenleft}eucl\ R{\isacharparenright}{\isacharparenright}{\isachardoublequoteclose}\ \isanewline
\isadelimproof
\ %
\endisadelimproof
\isatagproof
\isacommand{by}\isamarkupfalse%
\ {\isacharparenleft}metis\ eucl{\isacharunderscore}def\ refl{\isacharunderscore}def\ trans{\isacharunderscore}def{\isacharparenright}%
\endisatagproof
{\isafoldproof}%
\isadelimproof
\endisadelimproof
\isamarkupsubsubsection{M5 $\Longleftrightarrow$ M4B%
}
\isamarkuptrue%
\isacommand{theorem}\isamarkupfalse%
\ B{\isadigit{4}}{\isacharcolon}\ {\isachardoublequoteopen}{\isasymforall}R{\isachardot}{\isacharparenleft}{\isacharparenleft}refl\ R{\isacharparenright}\ {\isasymand}\ {\isacharparenleft}eucl\ R{\isacharparenright}{\isacharparenright}\ {\isasymlongleftrightarrow}\ {\isacharparenleft}{\isacharparenleft}refl\ R{\isacharparenright}\ {\isasymand}\ {\isacharparenleft}trans\ R{\isacharparenright}\ {\isasymand}\ {\isacharparenleft}sym\ R{\isacharparenright}{\isacharparenright}{\isachardoublequoteclose}\isanewline
\isadelimproof
\ %
\endisadelimproof
\isatagproof
\isacommand{by}\isamarkupfalse%
\ {\isacharparenleft}metis\ eucl{\isacharunderscore}def\ refl{\isacharunderscore}def\ sym{\isacharunderscore}def\ trans{\isacharunderscore}def{\isacharparenright}%
\endisatagproof
{\isafoldproof}%
\isadelimproof
\endisadelimproof
\isamarkupsubsubsection{M5 $\Longleftrightarrow$ D4B%
}
\isamarkuptrue%
\isacommand{theorem}\isamarkupfalse%
\ B{\isadigit{5}}{\isacharcolon}\ {\isachardoublequoteopen}{\isasymforall}R{\isachardot}{\isacharparenleft}{\isacharparenleft}refl\ R{\isacharparenright}\ {\isasymand}\ {\isacharparenleft}eucl\ R{\isacharparenright}{\isacharparenright}\ {\isasymlongleftrightarrow}\ {\isacharparenleft}{\isacharparenleft}ser\ R{\isacharparenright}\ {\isasymand}\ {\isacharparenleft}trans\ R{\isacharparenright}\ {\isasymand}\ {\isacharparenleft}sym\ R{\isacharparenright}{\isacharparenright}{\isachardoublequoteclose}\isanewline
\isadelimproof
\ %
\endisadelimproof
\isatagproof
\isacommand{by}\isamarkupfalse%
\ {\isacharparenleft}metis\ eucl{\isacharunderscore}def\ refl{\isacharunderscore}def\ ser{\isacharunderscore}def\ sym{\isacharunderscore}def\ trans{\isacharunderscore}def{\isacharparenright}%
\endisatagproof
{\isafoldproof}%
\isadelimproof
\endisadelimproof
\isamarkupsubsubsection{M5 $\Longleftrightarrow$ D4B5%
}
\isamarkuptrue%
\isacommand{theorem}\isamarkupfalse%
\ B{\isadigit{6}}{\isacharcolon}\ {\isachardoublequoteopen}{\isasymforall}R{\isachardot}{\isacharparenleft}{\isacharparenleft}refl\ R{\isacharparenright}\ {\isasymand}\ {\isacharparenleft}eucl\ R{\isacharparenright}{\isacharparenright}\ {\isasymlongleftrightarrow}\ {\isacharparenleft}{\isacharparenleft}ser\ R{\isacharparenright}\ {\isasymand}\ {\isacharparenleft}trans\ R{\isacharparenright}\ {\isasymand}\ {\isacharparenleft}sym\ R{\isacharparenright}\ {\isasymand}\ {\isacharparenleft}eucl\ R{\isacharparenright}{\isacharparenright}{\isachardoublequoteclose}\isanewline
\isadelimproof
\ %
\endisadelimproof
\isatagproof
\isacommand{by}\isamarkupfalse%
\ {\isacharparenleft}metis\ eucl{\isacharunderscore}def\ refl{\isacharunderscore}def\ ser{\isacharunderscore}def\ sym{\isacharunderscore}def\ trans{\isacharunderscore}def{\isacharparenright}%
\endisatagproof
{\isafoldproof}%
\isadelimproof
\endisadelimproof
\isamarkupsubsubsection{M5 $\Longleftrightarrow$ DB5%
}
\isamarkuptrue%
\isacommand{theorem}\isamarkupfalse%
\ B{\isadigit{7}}{\isacharcolon}\ {\isachardoublequoteopen}{\isasymforall}R{\isachardot}{\isacharparenleft}{\isacharparenleft}refl\ R{\isacharparenright}\ {\isasymand}\ {\isacharparenleft}eucl\ R{\isacharparenright}{\isacharparenright}\ {\isasymlongleftrightarrow}\ {\isacharparenleft}{\isacharparenleft}ser\ R{\isacharparenright}\ {\isasymand}\ {\isacharparenleft}sym\ R{\isacharparenright}\ {\isasymand}\ {\isacharparenleft}eucl\ R{\isacharparenright}{\isacharparenright}{\isachardoublequoteclose}\isanewline
\isadelimproof
\ %
\endisadelimproof
\isatagproof
\isacommand{by}\isamarkupfalse%
\ {\isacharparenleft}metis\ eucl{\isacharunderscore}def\ refl{\isacharunderscore}def\ ser{\isacharunderscore}def\ sym{\isacharunderscore}def{\isacharparenright}%
\endisatagproof
{\isafoldproof}%
\isadelimproof
\endisadelimproof
\isamarkupsubsubsection{KB5 $\Longleftrightarrow$ K4B5%
}
\isamarkuptrue%
\isacommand{theorem}\isamarkupfalse%
\ B{\isadigit{8}}{\isacharcolon}\ {\isachardoublequoteopen}{\isasymforall}R{\isachardot}{\isacharparenleft}{\isacharparenleft}sym\ R{\isacharparenright}\ {\isasymand}\ {\isacharparenleft}eucl\ R{\isacharparenright}{\isacharparenright}\ {\isasymlongleftrightarrow}\ {\isacharparenleft}{\isacharparenleft}trans\ R{\isacharparenright}\ {\isasymand}\ {\isacharparenleft}sym\ R{\isacharparenright}\ {\isasymand}\ {\isacharparenleft}eucl\ R{\isacharparenright}{\isacharparenright}{\isachardoublequoteclose}\isanewline
\isadelimproof
\ %
\endisadelimproof
\isatagproof
\isacommand{by}\isamarkupfalse%
\ {\isacharparenleft}metis\ eucl{\isacharunderscore}def\ sym{\isacharunderscore}def\ trans{\isacharunderscore}def{\isacharparenright}%
\endisatagproof
{\isafoldproof}%
\isadelimproof
\endisadelimproof
\isamarkupsubsubsection{KB5 $\Longleftrightarrow$ K4B%
}
\isamarkuptrue%
\isacommand{theorem}\isamarkupfalse%
\ B{\isadigit{9}}{\isacharcolon}\ {\isachardoublequoteopen}{\isasymforall}R{\isachardot}{\isacharparenleft}{\isacharparenleft}sym\ R{\isacharparenright}\ {\isasymand}\ {\isacharparenleft}eucl\ R{\isacharparenright}{\isacharparenright}\ {\isasymlongleftrightarrow}\ {\isacharparenleft}{\isacharparenleft}trans\ R{\isacharparenright}\ {\isasymand}\ {\isacharparenleft}sym\ R{\isacharparenright}{\isacharparenright}{\isachardoublequoteclose}\isanewline
\isadelimproof
\ %
\endisadelimproof
\isatagproof
\isacommand{by}\isamarkupfalse%
\ {\isacharparenleft}metis\ eucl{\isacharunderscore}def\ sym{\isacharunderscore}def\ trans{\isacharunderscore}def{\isacharparenright}%
\endisatagproof
{\isafoldproof}%
\isadelimproof
\endisadelimproof
\isamarkupsubsection{Proper Inclusion Relations between Different Modal Logics%
}
\isamarkuptrue%
\begin{isamarkuptext}%
An edge within the cube denotes an inclusion between the connected logics. In the forward direction, these can
be trivially shown valid through monotonicity of entailment and equivalence of the different
axiomatizations. For example, for the forward link from logic \textbf{K} to logic \textbf{B}, we need to show that every 
theorem of \textbf{K} is also a theorem of \textbf{B}; this simply means to
disregard the additional axiom B.
 Below, the crucial backward directions are proved. 
Informally, it is shown that through moving further
up in the cube (adding further axioms), theorems can be proved which were not provable before; this means
that the inclusions are proper.
We write $A > B$ to indicate that logic $A$ can prove strictly more theorems than logic $B$.

It has to be noted that some logics are actually equivalent if the only models considered have few
enough worlds; examples are given below. We introduce some useful abbreviations to formulate
constraints on the number of worlds in a model.%
\end{isamarkuptext}%
\isamarkuptrue%
\isacommand{abbreviation}\isamarkupfalse%
\ one{\isacharunderscore}world{\isacharunderscore}model\ {\isacharcolon}{\isacharcolon}\ {\isachardoublequoteopen}i\ {\isasymRightarrow}\ bool{\isachardoublequoteclose}\ \ \ \isakeyword{where}\ {\isachardoublequoteopen}{\isacharhash}\isactrlsup {\isadigit{1}}\ w{\isadigit{1}}\ {\isasymequiv}\ {\isasymforall}x{\isachardot}\ x\ {\isacharequal}\ w{\isadigit{1}}{\isachardoublequoteclose}\isanewline
\isacommand{abbreviation}\isamarkupfalse%
\ two{\isacharunderscore}world{\isacharunderscore}model\ {\isacharcolon}{\isacharcolon}\ {\isachardoublequoteopen}i\ {\isasymRightarrow}\ i\ {\isasymRightarrow}\ bool{\isachardoublequoteclose}\ \ \ \isakeyword{where}\ {\isachardoublequoteopen}{\isacharhash}\isactrlsup {\isadigit{2}}\ w{\isadigit{1}}\ w{\isadigit{2}}\ {\isasymequiv}\ {\isacharparenleft}{\isasymforall}x{\isachardot}\ x\ {\isacharequal}\ w{\isadigit{1}}\ {\isasymor}\ x\ {\isacharequal}\ w{\isadigit{2}}{\isacharparenright}\ {\isasymand}\ w{\isadigit{1}}\ {\isasymnoteq}\ w{\isadigit{2}}{\isachardoublequoteclose}\ \isanewline
\isacommand{abbreviation}\isamarkupfalse%
\ three{\isacharunderscore}world{\isacharunderscore}model\ {\isacharcolon}{\isacharcolon}\ {\isachardoublequoteopen}i\ {\isasymRightarrow}\ i\ {\isasymRightarrow}\ i\ {\isasymRightarrow}\ bool{\isachardoublequoteclose}\ \ \ \isakeyword{where}\ {\isachardoublequoteopen}{\isacharhash}\isactrlsup {\isadigit{3}}\ w{\isadigit{1}}\ w{\isadigit{2}}\ w{\isadigit{3}}\ {\isasymequiv}\ {\isacharparenleft}{\isasymforall}x{\isachardot}\ x\ {\isacharequal}\ w{\isadigit{1}}\ {\isasymor}\ x\ {\isacharequal}\ w{\isadigit{2}}\ {\isasymor}\ x\ {\isacharequal}\ w{\isadigit{3}}{\isacharparenright}\ {\isasymand}\ w{\isadigit{1}}\ {\isasymnoteq}\ w{\isadigit{2}}\ {\isasymand}\ w{\isadigit{1}}\ {\isasymnoteq}\ w{\isadigit{3}}\ {\isasymand}\ w{\isadigit{2}}\ {\isasymnoteq}\ w{\isadigit{3}}{\isachardoublequoteclose}%
\begin{isamarkuptext}%
In what follows, we reserve the symbols \emph{i1}, \emph{i2} and \emph{i3} for worlds, and \emph{r} for an accessibility relation.%
\end{isamarkuptext}%
\isamarkuptrue%
\begin{isamarkuptext}%
We applied the following methodology in the experiments reported in this section:

\begin{description}
\item[\textbf{(Step A)}] First we deliberately made invalid conjectures about inclusion relations---e.g. for proving 
K4 $>$ K we first wrongly conjectured that K4 $\subseteq$ K, meaning that K4 entails K. 
We did this by conjecturing 
\begin{center} \isa{lemma\ C{\isadigit{1}}{\isacharunderscore}A{\isacharcolon}\ {\isasymforall}R{\isachardot}\ {\isacharparenleft}trans\ R{\isacharparenright}} \end{center}
These wrongly-conjectured lemmata in Step A are uniformly named \isa{C{\isacharasterisk}{\isacharunderscore}A}.
Note that for the formulation of the \isa{C{\isacharasterisk}{\isacharunderscore}A}-lemmata we again exploit the correspondence results given earlier, 
and we work with conditions on the accessibility relations instead of using the corresponding modal logic axioms.
For each \isa{C{\isacharasterisk}{\isacharunderscore}A}-lemma Nitpick quickly generates a countermodel, which 
it communicates in a specific syntax. For example, the countermodel it presents for \isa{C{\isadigit{1}}{\isacharunderscore}A} is 
\begin{center}
\isa{\ R\ {\isacharequal}\ {\isacharparenleft}{\isasymlambda}x{\isachardot}\ {\isacharunderscore}{\isacharparenright}{\isacharparenleft}i\isactrlsub {\isadigit{1}}\ {\isacharcolon}{\isacharequal}\ {\isacharparenleft}{\isasymlambda}x{\isachardot}\ {\isacharunderscore}{\isacharparenright}{\isacharparenleft}i\isactrlsub {\isadigit{1}}\ {\isacharcolon}{\isacharequal}\ True{\isacharcomma}\ i\isactrlsub {\isadigit{2}}\ {\isacharcolon}{\isacharequal}\ True{\isacharparenright}{\isacharcomma}\ i\isactrlsub {\isadigit{2}}\ {\isacharcolon}{\isacharequal}\ {\isacharparenleft}{\isasymlambda}x{\isachardot}\ {\isacharunderscore}{\isacharparenright}{\isacharparenleft}i\isactrlsub {\isadigit{1}}\ {\isacharcolon}{\isacharequal}\ True{\isacharcomma}\ i\isactrlsub {\isadigit{2}}\ {\isacharcolon}{\isacharequal}\ False{\isacharparenright}{\isacharparenright}\ }.
\end{center}
Diagrammatically this 2-world countermodel can be represented as follows 
\begin{center}
\begin{tikzpicture}[shorten >=1pt,node distance=2cm,on grid,auto] 
   \node[state] (i_1)   {$i_1$}; 
   \node[state] (i_2) [right=of i_1] {$i_2$}; 
    \path[->] 
    (i_1) edge [loop above] node {} ()
          edge [bend left] node {} (i_2)
    (i_2) edge [bend left] node {} (i_1);
\end{tikzpicture}
\end{center}

\item[\textbf{(Step B)}] Next, we systematically employed the arity information obtained from the countermodels for the \isa{C{\isacharasterisk}{\isacharunderscore}A}-lemmata, reported by Nitpick,
to formulate a corresponding 
lemma to be passed via Sledgehammer to the HOL-ATPs LEO-II, Satallax and/or CVC4 \cite{CVC4} (whenever it was not trivially provable by the automation 
tools \isa{simp}, \isa{force} and/or \isa{blast} available within Isabelle/HOL).
In our running example this lemma is 
\begin{center}
 \isa{C{\isadigit{1}}{\isacharunderscore}B{\isacharcolon}\ {\isacharhash}\isactrlsup {\isadigit{2}}\ i{\isadigit{1}}\ i{\isadigit{2}}\ {\isasymlongrightarrow}\ {\isasymforall}R{\isachardot}\ {\isasymnot}\ trans\ R}
\end{center}
 All but four of these lemmata can actually be proved by either LEO-II or Satallax. Some of the easier problems can already be automated with 
 \isa{simp}, \isa{force} and  \isa{blast}, which are preferred here.  
 The four cases in which no automation attempts succeeded (we also tried all other integrated ATPs in Isabelle) 
 are named \isa{C{\isacharasterisk}{\isacharunderscore}ATP{\isacharunderscore}challenge} below.
 Moreover, there are ten problems named \isa{C{\isacharasterisk}{\isacharunderscore}Isabelle{\isacharunderscore}challenge}. For these problems LEO-II or Satallax found proofs, but their
 Metis-based integration into Isabelle failed. Hence, no verification was obtained for these problems. However, we found that 
 five of these \isa{C{\isacharasterisk}{\isacharunderscore}Isabelle{\isacharunderscore}challenge} problems can also be proved by CVC4, for which proof integration
 worked. 
 Unfortunately, no other automation means (including the integrated first-order ATPs or SMT solvers) succeeded for the
 \isa{C{\isacharasterisk}{\isacharunderscore}Isabelle{\isacharunderscore}challenge} problems.

\item[\textbf{(Step C)}] For the verification of the modal logic cube, the non-proved or non-integrated \isa{C{\isacharasterisk}{\isacharunderscore}challenge} problems of Step B are clearly unsatisfactory, since
no proper verification in Isabelle is obtained. However, an easy solution for these (and all other) cases
is possible by exploiting not only Nitpick's arity information on the countermodels, but by using all the information about the 
countermodels it presents, that is, the precise information on the accessibility relation. 
For example, Nitpick's countermodel for \isa{C{\isadigit{1}}{\isacharunderscore}A} from above 
can be converted into the following theorem
(where \isa{r} denotes a fixed accessibility relation) 
\begin{center}
\isa{theorem\ C{\isadigit{1}}{\isacharunderscore}C{\isacharcolon}\ {\isacharhash}\isactrlsup {\isadigit{2}}\ i{\isadigit{1}}\ i{\isadigit{2}}\ {\isasymand}\ r\ i{\isadigit{1}}\ i{\isadigit{1}}\ {\isasymand}\ r\ i{\isadigit{1}}\ i{\isadigit{2}}\ {\isasymand}\ r\ i{\isadigit{2}}\ i{\isadigit{1}}\ {\isasymand}\ {\isasymnot}r\ i{\isadigit{2}}\ i{\isadigit{2}}\ {\isasymlongrightarrow}\ {\isasymnot}\ trans\ r}. 
\end{center}
The resulting theorems we generate 
are uniformly named \isa{C{\isacharasterisk}{\isacharunderscore}C}. It turns out that all  \isa{C{\isacharasterisk}{\isacharunderscore}C}-theorems can be quickly verified in Isabelle by Metis. 
Thus, for each link in the modal logic we provide either a verified \isa{C{\isacharasterisk}{\isacharunderscore}B} theorem or, if this was not successful, a verified \isa{C{\isacharasterisk}{\isacharunderscore}C} 
theorem. Taken together, this confirms that the inclusion relation in the cube are indeed proper. 
Hence, these \isa{C{\isacharasterisk}{\isacharunderscore}B} resp. \isa{C{\isacharasterisk}{\isacharunderscore}C} theorems complete the verification of the modal logic cube. Below the 
\isa{C{\isacharasterisk}{\isacharunderscore}C} proof attempts are omitted if the corresponding \isa{C{\isacharasterisk}{\isacharunderscore}B} attempts were already successful.

\item[\textbf{(Step D)}] We additionally prove that the countermodels found by Nitpick in Step A are minimal (regarding the number 
of possible worlds). In other words, we prove here that the world model constraints as exploited in Step B are in fact minimal constraints
under which the inclusion relations can be shown to be proper. Of course, if such a countermodel consists of one possible world only, nothing 
needs to be shown. 
\end{description}

Note that the entire process sketched above, that is the schematic Steps A-D, could be fully automated, meaning that the formulation of the lemmata and theorems
in each step could be obtained automatically by analyzing and converting Nitpick's output.
In our experiments we still wrote and invoked the verification of each link in the 
modal cube manually however. Clearly, automation facilities could be very useful for the exploration of the meta-theory of other logics, for
example, conditional logics~\cite{IJCAI}, since the overall methodology is obviously transferable to other logics of interest.%
\end{isamarkuptext}%
\isamarkuptrue%
\begin{isamarkuptext}%
\begin{isbfig}{7em}
\begin{tikzpicture}[shorten >=1pt,node distance=2cm,on grid,auto] 
   \node[state] (i_1)   {$i_1$}; 
   \node[state] (i_2) [right=of i_1] {$i_2$}; 
    \path[->] 
    (i_1) edge [loop above] node {} ()
          edge [bend left] node {} (i_2)
    (i_2) edge [bend left] node {} (i_1);
\end{tikzpicture}
\end{isbfig}%
\end{isamarkuptext}%
\isamarkuptrue%
\isamarkupsubsubsection{K4 $>$ K%
}
\isamarkuptrue%
\isacommand{lemma}\isamarkupfalse%
\ C{\isadigit{1}}{\isacharunderscore}A{\isacharcolon}\ {\isachardoublequoteopen}{\isasymforall}R{\isachardot}\ trans\ R{\isachardoublequoteclose}\ \isacommand{nitpick}\isamarkupfalse%
\isadelimproof
\ %
\endisadelimproof
\isatagproof
\isacommand{oops}\isamarkupfalse%
\endisatagproof
{\isafoldproof}%
\isadelimproof
\endisadelimproof
\isanewline
\isacommand{theorem}\isamarkupfalse%
\ C{\isadigit{1}}{\isacharunderscore}B{\isacharcolon}\ {\isachardoublequoteopen}{\isacharhash}\isactrlsup {\isadigit{2}}\ i{\isadigit{1}}\ i{\isadigit{2}}\ {\isasymlongrightarrow}\ {\isasymnot}\ {\isacharparenleft}{\isasymforall}R{\isachardot}\ trans\ R{\isacharparenright}{\isachardoublequoteclose}%
\isadelimproof
\ %
\endisadelimproof
\isatagproof
\isacommand{by}\isamarkupfalse%
\ {\isacharparenleft}simp\ add{\isacharcolon}trans{\isacharunderscore}def{\isacharcomma}\ force{\isacharparenright}%
\endisatagproof
{\isafoldproof}%
\isadelimproof
\endisadelimproof
\isanewline
\isacommand{lemma}\isamarkupfalse%
\ C{\isadigit{1}}{\isacharunderscore}D{\isacharcolon}\ {\isachardoublequoteopen}{\isacharhash}\isactrlsup {\isadigit{1}}\ i{\isadigit{1}}\ {\isasymlongrightarrow}\ {\isacharparenleft}{\isasymforall}R{\isachardot}\ trans\ R{\isacharparenright}{\isachardoublequoteclose}%
\isadelimproof
\ %
\endisadelimproof
\isatagproof
\isacommand{by}\isamarkupfalse%
\ {\isacharparenleft}metis\ {\isacharparenleft}lifting{\isacharcomma}\ full{\isacharunderscore}types{\isacharparenright}\ trans{\isacharunderscore}def{\isacharparenright}%
\endisatagproof
{\isafoldproof}%
\isadelimproof
\endisadelimproof
\begin{isamarkuptext}%
\begin{isbfig}{7em}
\begin{tikzpicture}[shorten >=1pt,node distance=2cm,on grid,auto] 
   \node[state] (i_1)   {$i_1$}; 
   \node[state] (i_2) [right=of i_1] {$i_2$}; 
    \path[->] 
    (i_2) edge [loop above] node {} ()
          edge node {} (i_1);
\end{tikzpicture}
\end{isbfig}%
\end{isamarkuptext}%
\isamarkuptrue%
\isamarkupsubsubsection{K5 $>$ K%
}
\isamarkuptrue%
\isacommand{lemma}\isamarkupfalse%
\ C{\isadigit{2}}{\isacharunderscore}A{\isacharcolon}\ {\isachardoublequoteopen}{\isasymforall}R{\isachardot}\ eucl\ R{\isachardoublequoteclose}\ \isacommand{nitpick}\isamarkupfalse%
\isadelimproof
\ %
\endisadelimproof
\isatagproof
\isacommand{oops}\isamarkupfalse%
\endisatagproof
{\isafoldproof}%
\isadelimproof
\endisadelimproof
\isanewline
\isacommand{theorem}\isamarkupfalse%
\ C{\isadigit{2}}{\isacharunderscore}B{\isacharcolon}\ {\isachardoublequoteopen}{\isacharhash}\isactrlsup {\isadigit{2}}\ i{\isadigit{1}}\ i{\isadigit{2}}\ {\isasymlongrightarrow}\ {\isasymnot}\ {\isacharparenleft}{\isasymforall}R{\isachardot}\ eucl\ R{\isacharparenright}{\isachardoublequoteclose}%
\isadelimproof
\ %
\endisadelimproof
\isatagproof
\isacommand{by}\isamarkupfalse%
\ {\isacharparenleft}simp\ add{\isacharcolon}eucl{\isacharunderscore}def{\isacharcomma}\ force{\isacharparenright}%
\endisatagproof
{\isafoldproof}%
\isadelimproof
\endisadelimproof
\isanewline
\isacommand{lemma}\isamarkupfalse%
\ C{\isadigit{2}}{\isacharunderscore}D{\isacharcolon}\ {\isachardoublequoteopen}{\isacharhash}\isactrlsup {\isadigit{1}}\ i{\isadigit{1}}\ {\isasymlongrightarrow}\ {\isacharparenleft}{\isasymforall}R{\isachardot}\ eucl\ R{\isacharparenright}{\isachardoublequoteclose}%
\isadelimproof
\ %
\endisadelimproof
\isatagproof
\isacommand{by}\isamarkupfalse%
\ {\isacharparenleft}metis\ {\isacharparenleft}lifting{\isacharcomma}\ full{\isacharunderscore}types{\isacharparenright}\ eucl{\isacharunderscore}def{\isacharparenright}%
\endisatagproof
{\isafoldproof}%
\isadelimproof
\endisadelimproof
\begin{isamarkuptext}%
\begin{isbfig}{7em}
\begin{tikzpicture}[shorten >=1pt,node distance=2cm,on grid,auto] 
   \node[state] (i_1)   {$i_1$}; 
   \node[state] (i_2) [right=of i_1] {$i_2$}; 
    \path[->] 
    (i_2) edge node {} (i_1);
\end{tikzpicture}
\end{isbfig}%
\end{isamarkuptext}%
\isamarkuptrue%
\isamarkupsubsubsection{KB $>$ K%
}
\isamarkuptrue%
\isacommand{lemma}\isamarkupfalse%
\ C{\isadigit{3}}{\isacharunderscore}A{\isacharcolon}\ {\isachardoublequoteopen}{\isasymforall}R{\isachardot}\ sym\ R{\isachardoublequoteclose}\ \isacommand{nitpick}\isamarkupfalse%
\isadelimproof
\ %
\endisadelimproof
\isatagproof
\isacommand{oops}\isamarkupfalse%
\endisatagproof
{\isafoldproof}%
\isadelimproof
\endisadelimproof
\isanewline
\isacommand{theorem}\isamarkupfalse%
\ C{\isadigit{3}}{\isacharunderscore}B{\isacharcolon}\ {\isachardoublequoteopen}{\isacharhash}\isactrlsup {\isadigit{2}}\ i{\isadigit{1}}\ i{\isadigit{2}}\ {\isasymlongrightarrow}\ {\isasymnot}\ {\isacharparenleft}{\isasymforall}R{\isachardot}\ sym\ R{\isacharparenright}{\isachardoublequoteclose}%
\isadelimproof
\ %
\endisadelimproof
\isatagproof
\isacommand{by}\isamarkupfalse%
\ {\isacharparenleft}simp\ add{\isacharcolon}sym{\isacharunderscore}def{\isacharcomma}\ force{\isacharparenright}%
\endisatagproof
{\isafoldproof}%
\isadelimproof
\endisadelimproof
\isanewline
\isacommand{lemma}\isamarkupfalse%
\ C{\isadigit{3}}{\isacharunderscore}D{\isacharcolon}\ {\isachardoublequoteopen}{\isacharhash}\isactrlsup {\isadigit{1}}\ i{\isadigit{1}}\ {\isasymlongrightarrow}\ {\isacharparenleft}{\isasymforall}R{\isachardot}\ sym\ R{\isacharparenright}{\isachardoublequoteclose}%
\isadelimproof
\ %
\endisadelimproof
\isatagproof
\isacommand{by}\isamarkupfalse%
\ {\isacharparenleft}metis\ {\isacharparenleft}full{\isacharunderscore}types{\isacharparenright}\ sym{\isacharunderscore}def{\isacharparenright}%
\endisatagproof
{\isafoldproof}%
\isadelimproof
\endisadelimproof
\isamarkupsubsubsection{K45 $>$ K4%
}
\isamarkuptrue%
\begin{isamarkuptext}%
\begin{isbfig}{7em}
\begin{tikzpicture}[shorten >=1pt,node distance=2cm,on grid,auto] 
   \node[state] (i_1)   {$i_1$}; 
   \node[state] (i_2) [right=of i_1] {$i_2$}; 
    \path[->] 
    (i_1) edge [bend left] node {} (i_2)
    (i_2) edge [bend left] node {} (i_1);
\end{tikzpicture}
\end{isbfig}%
\end{isamarkuptext}%
\isamarkuptrue%
\isacommand{lemma}\isamarkupfalse%
\ C{\isadigit{4}}{\isacharunderscore}A{\isacharcolon}\ {\isachardoublequoteopen}{\isasymforall}R{\isachardot}\ ser\ R\ {\isasymlongrightarrow}\ {\isacharparenleft}ser\ R\ {\isasymand}\ eucl\ R{\isacharparenright}{\isachardoublequoteclose}\ \isacommand{nitpick}\isamarkupfalse%
\isadelimproof
\ %
\endisadelimproof
\isatagproof
\isacommand{oops}\isamarkupfalse%
\endisatagproof
{\isafoldproof}%
\isadelimproof
\endisadelimproof
\isanewline
\isacommand{lemma}\isamarkupfalse%
\ C{\isadigit{4}}{\isacharunderscore}B{\isacharunderscore}Isabelle{\isacharunderscore}challenge{\isacharcolon}\ {\isachardoublequoteopen}{\isacharhash}\isactrlsup {\isadigit{2}}\ i{\isadigit{1}}\ i{\isadigit{2}}\ {\isasymlongrightarrow}\ {\isasymnot}\ {\isacharparenleft}{\isasymforall}R{\isachardot}\ ser\ R\ {\isasymlongrightarrow}\ {\isacharparenleft}ser\ R\ {\isasymand}\ eucl\ R{\isacharparenright}{\isacharparenright}{\isachardoublequoteclose}\ \ \isanewline
\isamarkupcmt{sledgehammer [remote\_leo2](ser\_def eucl\_def)%
}
\isanewline
\isamarkupcmt{CPU time: 13.74 s. Metis reconstruction failed.%
}
\isanewline
\isamarkupcmt{sledgehammer [cvc4,timeout=300] -- timed out%
}
\isadelimproof
\ %
\endisadelimproof
\isatagproof
\isacommand{oops}\isamarkupfalse%
\endisatagproof
{\isafoldproof}%
\isadelimproof
\endisadelimproof
\isanewline
\isacommand{theorem}\isamarkupfalse%
\ C{\isadigit{4}}{\isacharunderscore}C{\isacharcolon}\ {\isachardoublequoteopen}{\isacharhash}\isactrlsup {\isadigit{2}}\ i{\isadigit{1}}\ i{\isadigit{2}}\ \ {\isasymand}\ {\isasymnot}r\ i{\isadigit{1}}\ i{\isadigit{1}}\ {\isasymand}\ r\ i{\isadigit{1}}\ i{\isadigit{2}}\ {\isasymand}\ r\ i{\isadigit{2}}\ i{\isadigit{1}}\ {\isasymand}\ {\isasymnot}r\ i{\isadigit{2}}\ i{\isadigit{2}}\ {\isasymlongrightarrow}\ {\isasymnot}\ {\isacharparenleft}ser\ r\ {\isasymlongrightarrow}\ {\isacharparenleft}ser\ r\ {\isasymand}\ eucl\ r{\isacharparenright}{\isacharparenright}{\isachardoublequoteclose}\ \isanewline
\isadelimproof
\ %
\endisadelimproof
\isatagproof
\isacommand{by}\isamarkupfalse%
\ {\isacharparenleft}metis\ ser{\isacharunderscore}def\ eucl{\isacharunderscore}def{\isacharparenright}%
\endisatagproof
{\isafoldproof}%
\isadelimproof
\isanewline
\endisadelimproof
\isacommand{lemma}\isamarkupfalse%
\ C{\isadigit{4}}{\isacharunderscore}D{\isacharcolon}\ {\isachardoublequoteopen}{\isacharhash}\isactrlsup {\isadigit{1}}\ i{\isadigit{1}}\ \ {\isasymlongrightarrow}\ {\isacharparenleft}{\isasymforall}R{\isachardot}\ ser\ R\ {\isasymlongrightarrow}\ {\isacharparenleft}ser\ R\ {\isasymand}\ eucl\ R{\isacharparenright}{\isacharparenright}{\isachardoublequoteclose}%
\isadelimproof
\ %
\endisadelimproof
\isatagproof
\isacommand{by}\isamarkupfalse%
\ {\isacharparenleft}metis\ {\isacharparenleft}full{\isacharunderscore}types{\isacharparenright}\ eucl{\isacharunderscore}def{\isacharparenright}%
\endisatagproof
{\isafoldproof}%
\isadelimproof
\endisadelimproof
\begin{isamarkuptext}%
\begin{isbfig}{7em}
\begin{tikzpicture}[shorten >=1pt,node distance=2cm,on grid,auto] 
   \node[state] (i_1)   {$i_1$};
\end{tikzpicture}
\end{isbfig}%
\end{isamarkuptext}%
\isamarkuptrue%
\isamarkupsubsubsection{K45 $>$ K5%
}
\isamarkuptrue%
\isacommand{lemma}\isamarkupfalse%
\ C{\isadigit{5}}{\isacharunderscore}A{\isacharcolon}\ {\isachardoublequoteopen}{\isasymforall}R{\isachardot}\ eucl\ R\ {\isasymlongrightarrow}\ {\isacharparenleft}ser\ R\ {\isasymand}\ eucl\ R{\isacharparenright}{\isachardoublequoteclose}\ \isanewline
\ \isacommand{nitpick}\isamarkupfalse%
\isadelimproof
\ %
\endisadelimproof
\isatagproof
\isacommand{oops}\isamarkupfalse%
\endisatagproof
{\isafoldproof}%
\isadelimproof
\endisadelimproof
\isanewline
\isacommand{lemma}\isamarkupfalse%
\ C{\isadigit{5}}{\isacharunderscore}B{\isacharunderscore}Isabelle{\isacharunderscore}challenge{\isacharcolon}\ {\isachardoublequoteopen}{\isacharhash}\isactrlsup {\isadigit{1}}\ i{\isadigit{1}}\ {\isasymlongrightarrow}\ {\isasymnot}\ {\isacharparenleft}{\isasymforall}R{\isachardot}\ {\isacharparenleft}eucl\ R{\isacharparenright}\ {\isasymlongrightarrow}\ {\isacharparenleft}ser\ R{\isacharparenright}\ {\isasymand}\ {\isacharparenleft}eucl\ R{\isacharparenright}{\isacharparenright}{\isachardoublequoteclose}\ \isanewline
\isamarkupcmt{sledgehammer [remote\_leo2](eucl\_def ser\_def) -- CPU time: 14.61 s. Metis reconstruction failed.%
}
\ \isanewline
\isamarkupcmt{sledgehammer [cvc4,timeout=300] -- timed out%
}
\isadelimproof
\ %
\endisadelimproof
\isatagproof
\isacommand{oops}\isamarkupfalse%
\endisatagproof
{\isafoldproof}%
\isadelimproof
\endisadelimproof
\isanewline
\isacommand{theorem}\isamarkupfalse%
\ C{\isadigit{5}}{\isacharunderscore}C{\isacharcolon}\ {\isachardoublequoteopen}{\isacharhash}\isactrlsup {\isadigit{1}}\ i{\isadigit{1}}\ {\isasymand}\ {\isasymnot}r\ i{\isadigit{1}}\ i{\isadigit{1}}\ {\isasymlongrightarrow}\ {\isasymnot}\ {\isacharparenleft}eucl\ r\ {\isasymlongrightarrow}\ {\isacharparenleft}ser\ r\ {\isasymand}\ eucl\ r{\isacharparenright}{\isacharparenright}{\isachardoublequoteclose}%
\isadelimproof
\ %
\endisadelimproof
\isatagproof
\isacommand{by}\isamarkupfalse%
\ {\isacharparenleft}metis\ {\isacharparenleft}full{\isacharunderscore}types{\isacharparenright}\ eucl{\isacharunderscore}def\ ser{\isacharunderscore}def{\isacharparenright}%
\endisatagproof
{\isafoldproof}%
\isadelimproof
\endisadelimproof
\begin{isamarkuptext}%
\begin{isbfig}{7em}
\begin{tikzpicture}[shorten >=1pt,node distance=2cm,on grid,auto] 
   \node[state] (i_1)   {$i_1$}; 
   \node[state] (i_2) [right=of i_1] {$i_2$}; 
    \path[->] 
    (i_1) edge [loop above] node {} () 
          edge [bend left] node {} (i_2) 
    (i_2) edge [bend left] node {} (i_1);
\end{tikzpicture}
\end{isbfig}%
\end{isamarkuptext}%
\isamarkuptrue%
\isamarkupsubsubsection{KB5 $>$ KB%
}
\isamarkuptrue%
\isacommand{lemma}\isamarkupfalse%
\ C{\isadigit{6}}{\isacharunderscore}A{\isacharcolon}\ {\isachardoublequoteopen}{\isasymforall}R{\isachardot}\ sym\ R\ {\isasymlongrightarrow}\ {\isacharparenleft}sym\ R\ {\isasymand}\ eucl\ R{\isacharparenright}{\isachardoublequoteclose}\ \isanewline
\ \isacommand{nitpick}\isamarkupfalse%
\isadelimproof
\ %
\endisadelimproof
\isatagproof
\isacommand{oops}\isamarkupfalse%
\endisatagproof
{\isafoldproof}%
\isadelimproof
\endisadelimproof
\isanewline
\isacommand{lemma}\isamarkupfalse%
\ C{\isadigit{6}}{\isacharunderscore}B{\isacharunderscore}Isabelle{\isacharunderscore}challenge{\isacharcolon}\ {\isachardoublequoteopen}{\isacharhash}\isactrlsup {\isadigit{2}}\ i{\isadigit{1}}\ i{\isadigit{2}}\ {\isasymlongrightarrow}\ {\isasymnot}\ {\isacharparenleft}{\isasymforall}R{\isachardot}\ sym\ R\ {\isasymlongrightarrow}\ {\isacharparenleft}sym\ R\ {\isasymand}\ eucl\ R{\isacharparenright}{\isacharparenright}{\isachardoublequoteclose}\isanewline
\isamarkupcmt{sledgehammer [remote\_leo2,timeout=200](sym\_def eucl\_def) -- CPU time: 29.0 s. Metis reconstruction failed.%
}
\isanewline
\isamarkupcmt{sledgehammer [cvc4,timeout=300] suggested following line:%
}
\isanewline
\isadelimproof
\ %
\endisadelimproof
\isatagproof
\isacommand{by}\isamarkupfalse%
\ {\isacharparenleft}metis\ {\isacharparenleft}full{\isacharunderscore}types{\isacharparenright}\ A{\isadigit{4}}\ B{\isadigit{8}}\ C{\isadigit{1}}{\isacharunderscore}B\ IV{\isacharunderscore}def\ sym{\isacharunderscore}def{\isacharparenright}%
\endisatagproof
{\isafoldproof}%
\isadelimproof
\endisadelimproof
\ \isanewline
\isacommand{lemma}\isamarkupfalse%
\ C{\isadigit{6}}{\isacharunderscore}D{\isacharcolon}\ {\isachardoublequoteopen}{\isacharhash}\isactrlsup {\isadigit{1}}\ i{\isadigit{1}}\ {\isasymlongrightarrow}\ {\isacharparenleft}{\isasymforall}R{\isachardot}\ sym\ R\ {\isasymlongrightarrow}\ {\isacharparenleft}sym\ R\ {\isasymand}\ eucl\ R{\isacharparenright}{\isacharparenright}{\isachardoublequoteclose}\isanewline
\isadelimproof
\ %
\endisadelimproof
\isatagproof
\isacommand{by}\isamarkupfalse%
\ {\isacharparenleft}metis\ {\isacharparenleft}full{\isacharunderscore}types{\isacharparenright}\ eucl{\isacharunderscore}def{\isacharparenright}%
\endisatagproof
{\isafoldproof}%
\isadelimproof
\endisadelimproof
\begin{isamarkuptext}%
\begin{isbfig}{8em}
\begin{tikzpicture}[shorten >=1pt,node distance=2cm,on grid,auto] 
   \node[state] (i_1)   {$i_1$}; 
   \node[state] (i_2) [right=of i_1] {$i_2$}; 
    \path[->] 
    (i_1) edge [loop above] node {} ()
    (i_2) edge node {} (i_1);
\end{tikzpicture}
\end{isbfig}%
\end{isamarkuptext}%
\isamarkuptrue%
\isamarkupsubsubsection{KB5 $>$ K45%
}
\isamarkuptrue%
\isacommand{lemma}\isamarkupfalse%
\ C{\isadigit{7}}{\isacharunderscore}A{\isacharcolon}\ {\isachardoublequoteopen}{\isasymforall}R{\isachardot}\ ser\ R\ {\isasymand}\ eucl\ R\ {\isasymlongrightarrow}\ {\isacharparenleft}sym\ R\ {\isasymand}\ eucl\ R{\isacharparenright}{\isachardoublequoteclose}\ \isanewline
\ \isacommand{nitpick}\isamarkupfalse%
\isadelimproof
\ %
\endisadelimproof
\isatagproof
\isacommand{oops}\isamarkupfalse%
\endisatagproof
{\isafoldproof}%
\isadelimproof
\endisadelimproof
\isanewline
\isacommand{lemma}\isamarkupfalse%
\ C{\isadigit{7}}{\isacharunderscore}B{\isacharunderscore}Isabelle{\isacharunderscore}challenge{\isacharcolon}\ {\isachardoublequoteopen}{\isacharhash}\isactrlsup {\isadigit{2}}\ i{\isadigit{1}}\ i{\isadigit{2}}\ {\isasymlongrightarrow}\ {\isasymnot}\ {\isacharparenleft}{\isasymforall}R{\isachardot}\ ser\ R\ {\isasymand}\ eucl\ R\ {\isasymlongrightarrow}\ {\isacharparenleft}sym\ R\ {\isasymand}\ eucl\ R{\isacharparenright}{\isacharparenright}{\isachardoublequoteclose}\isanewline
\isamarkupcmt{sledgehammer [remote\_leo2] (ser\_def eucl\_def sym\_def) -- CPU time: 11.15 s. Metis reconstruction failed.%
}
\ \isanewline
\isamarkupcmt{sledgehammer [cvc4,timeout=300] -- timed out%
}
\isadelimproof
\ %
\endisadelimproof
\isatagproof
\isacommand{oops}\isamarkupfalse%
\endisatagproof
{\isafoldproof}%
\isadelimproof
\endisadelimproof
\ \isanewline
\isacommand{theorem}\isamarkupfalse%
\ C{\isadigit{7}}{\isacharunderscore}C{\isacharcolon}\ {\isachardoublequoteopen}{\isacharhash}\isactrlsup {\isadigit{2}}\ i{\isadigit{1}}\ i{\isadigit{2}}\ {\isasymand}\ r\ i{\isadigit{1}}\ i{\isadigit{1}}\ {\isasymand}\ {\isasymnot}\ r\ i{\isadigit{1}}\ i{\isadigit{2}}\ {\isasymand}\ r\ i{\isadigit{2}}\ i{\isadigit{1}}\ {\isasymand}\ {\isasymnot}\ r\ i{\isadigit{2}}\ i{\isadigit{2}}\ {\isasymlongrightarrow}\ {\isasymnot}\ {\isacharparenleft}ser\ r\ {\isasymand}\ eucl\ r\ {\isasymlongrightarrow}\ {\isacharparenleft}sym\ r\ {\isasymand}\ eucl\ r{\isacharparenright}{\isacharparenright}{\isachardoublequoteclose}\ \isanewline
\isadelimproof
\ %
\endisadelimproof
\isatagproof
\isacommand{by}\isamarkupfalse%
\ {\isacharparenleft}metis\ {\isacharparenleft}full{\isacharunderscore}types{\isacharparenright}\ ser{\isacharunderscore}def\ eucl{\isacharunderscore}def\ sym{\isacharunderscore}def{\isacharparenright}%
\endisatagproof
{\isafoldproof}%
\isadelimproof
\isanewline
\endisadelimproof
\isacommand{lemma}\isamarkupfalse%
\ C{\isadigit{7}}{\isacharunderscore}D{\isacharcolon}\ {\isachardoublequoteopen}{\isacharhash}\isactrlsup {\isadigit{1}}\ i{\isadigit{1}}\ {\isasymlongrightarrow}\ {\isacharparenleft}{\isasymforall}R{\isachardot}\ ser\ R\ {\isasymand}\ eucl\ R\ {\isasymlongrightarrow}\ {\isacharparenleft}sym\ R\ {\isasymand}\ eucl\ R{\isacharparenright}{\isacharparenright}{\isachardoublequoteclose}%
\isadelimproof
\ %
\endisadelimproof
\isatagproof
\isacommand{by}\isamarkupfalse%
\ {\isacharparenleft}metis\ {\isacharparenleft}full{\isacharunderscore}types{\isacharparenright}\ sym{\isacharunderscore}def{\isacharparenright}%
\endisatagproof
{\isafoldproof}%
\isadelimproof
\endisadelimproof
\begin{isamarkuptext}%
\begin{isbfig}{7em}
\begin{tikzpicture}[shorten >=1pt,node distance=2cm,on grid,auto] 
   \node[state] (i_1)   {$i_1$}; 
\end{tikzpicture}
\end{isbfig}%
\end{isamarkuptext}%
\isamarkuptrue%
\isamarkupsubsubsection{D $>$ K%
}
\isamarkuptrue%
\isacommand{lemma}\isamarkupfalse%
\ C{\isadigit{8}}{\isacharunderscore}A{\isacharcolon}\ {\isachardoublequoteopen}{\isasymforall}R{\isachardot}\ ser\ R{\isachardoublequoteclose}\ \isacommand{nitpick}\isamarkupfalse%
\isadelimproof
\ %
\endisadelimproof
\isatagproof
\isacommand{oops}\isamarkupfalse%
\endisatagproof
{\isafoldproof}%
\isadelimproof
\endisadelimproof
\isanewline
\isacommand{lemma}\isamarkupfalse%
\ C{\isadigit{8}}{\isacharunderscore}B{\isacharcolon}\ {\isachardoublequoteopen}{\isacharhash}\isactrlsup {\isadigit{1}}\ i{\isadigit{1}}\ {\isasymlongrightarrow}\ {\isasymnot}{\isacharparenleft}{\isasymforall}R{\isachardot}\ {\isacharparenleft}ser\ R{\isacharparenright}{\isacharparenright}{\isachardoublequoteclose}%
\isadelimproof
\ %
\endisadelimproof
\isatagproof
\isacommand{by}\isamarkupfalse%
\ {\isacharparenleft}simp\ add{\isacharcolon}ser{\isacharunderscore}def{\isacharcomma}\ force{\isacharparenright}%
\endisatagproof
{\isafoldproof}%
\isadelimproof
\endisadelimproof
\isanewline
\isacommand{theorem}\isamarkupfalse%
\ C{\isadigit{8}}{\isacharunderscore}C{\isacharcolon}\ {\isachardoublequoteopen}{\isacharhash}\isactrlsup {\isadigit{1}}\ i{\isadigit{1}}\ {\isasymand}\ {\isasymnot}r\ i{\isadigit{1}}\ i{\isadigit{1}}\ {\isasymlongrightarrow}\ {\isasymnot}{\isacharparenleft}ser\ r{\isacharparenright}{\isachardoublequoteclose}%
\isadelimproof
\ %
\endisadelimproof
\isatagproof
\isacommand{by}\isamarkupfalse%
\ {\isacharparenleft}metis\ {\isacharparenleft}full{\isacharunderscore}types{\isacharparenright}\ ser{\isacharunderscore}def{\isacharparenright}%
\endisatagproof
{\isafoldproof}%
\isadelimproof
\endisadelimproof
\begin{isamarkuptext}%
\begin{isbfig}{7em}
\begin{tikzpicture}[shorten >=1pt,node distance=2cm,on grid,auto] 
   \node[state] (i_1)   {$i_1$}; 
\end{tikzpicture}
\end{isbfig}%
\end{isamarkuptext}%
\isamarkuptrue%
\isamarkupsubsubsection{D4 $>$ K4%
}
\isamarkuptrue%
\isacommand{lemma}\isamarkupfalse%
\ C{\isadigit{9}}{\isacharunderscore}A{\isacharcolon}\ {\isachardoublequoteopen}{\isasymforall}R{\isachardot}\ trans\ R\ {\isasymlongrightarrow}\ {\isacharparenleft}ser\ R\ {\isasymand}\ trans\ R{\isacharparenright}{\isachardoublequoteclose}\ \isanewline
\ \isacommand{nitpick}\isamarkupfalse%
\isadelimproof
\ %
\endisadelimproof
\isatagproof
\isacommand{oops}\isamarkupfalse%
\endisatagproof
{\isafoldproof}%
\isadelimproof
\endisadelimproof
\isanewline
\isacommand{theorem}\isamarkupfalse%
\ C{\isadigit{9}}{\isacharunderscore}B{\isacharcolon}\ {\isachardoublequoteopen}{\isacharhash}\isactrlsup {\isadigit{1}}\ i{\isadigit{1}}\ {\isasymlongrightarrow}\ {\isasymnot}\ {\isacharparenleft}{\isasymforall}R{\isachardot}\ trans\ R\ {\isasymlongrightarrow}\ {\isacharparenleft}ser\ R\ {\isasymand}\ trans\ R{\isacharparenright}{\isacharparenright}{\isachardoublequoteclose}\ \isanewline
\isadelimproof
\ %
\endisadelimproof
\isatagproof
\isacommand{using}\isamarkupfalse%
\ C{\isadigit{1}}{\isacharunderscore}D\ C{\isadigit{8}}{\isacharunderscore}B\ \isacommand{by}\isamarkupfalse%
\ blast%
\endisatagproof
{\isafoldproof}%
\isadelimproof
\endisadelimproof
\begin{isamarkuptext}%
\begin{isbfig}{7em}
\begin{tikzpicture}[shorten >=1pt,node distance=2cm,on grid,auto] 
   \node[state] (i_1)   {$i_1$}; 
\end{tikzpicture}
\end{isbfig}%
\end{isamarkuptext}%
\isamarkuptrue%
\isamarkupsubsubsection{D5 $>$ K5%
}
\isamarkuptrue%
\isacommand{lemma}\isamarkupfalse%
\ C{\isadigit{1}}{\isadigit{0}}{\isacharunderscore}A{\isacharcolon}\ {\isachardoublequoteopen}{\isasymforall}R{\isachardot}\ eucl\ R\ {\isasymlongrightarrow}\ {\isacharparenleft}ser\ R\ {\isasymand}\ eucl\ R{\isacharparenright}{\isachardoublequoteclose}\ \isacommand{nitpick}\isamarkupfalse%
\isadelimproof
\ %
\endisadelimproof
\isatagproof
\isacommand{oops}\isamarkupfalse%
\endisatagproof
{\isafoldproof}%
\isadelimproof
\endisadelimproof
\isanewline
\isacommand{theorem}\isamarkupfalse%
\ C{\isadigit{1}}{\isadigit{0}}{\isacharunderscore}B{\isacharcolon}\ {\isachardoublequoteopen}{\isacharhash}\isactrlsup {\isadigit{1}}\ i{\isadigit{1}}\ \ {\isasymlongrightarrow}\ {\isasymnot}\ {\isacharparenleft}{\isasymforall}R{\isachardot}\ eucl\ R\ {\isasymlongrightarrow}\ {\isacharparenleft}ser\ R\ {\isasymand}\ eucl\ R{\isacharparenright}{\isacharparenright}{\isachardoublequoteclose}%
\isadelimproof
\ %
\endisadelimproof
\isatagproof
\isacommand{using}\isamarkupfalse%
\ B{\isadigit{9}}\ C{\isadigit{3}}{\isacharunderscore}D\ C{\isadigit{9}}{\isacharunderscore}B\ \isacommand{by}\isamarkupfalse%
\ blast%
\endisatagproof
{\isafoldproof}%
\isadelimproof
\endisadelimproof
\begin{isamarkuptext}%
\begin{isbfig}{7em}
\begin{tikzpicture}[shorten >=1pt,node distance=2cm,on grid,auto] 
   \node[state] (i_1)   {$i_1$}; 
   \node[state] (i_2) [right=of i_1] {$i_2$}; 
    \path[->] 
    (i_1) edge [loop above] node {} ();
\end{tikzpicture}
\end{isbfig}%
\end{isamarkuptext}%
\isamarkuptrue%
\isamarkupsubsubsection{D45 $>$ K45%
}
\isamarkuptrue%
\isacommand{lemma}\isamarkupfalse%
\ C{\isadigit{1}}{\isadigit{1}}{\isacharunderscore}A{\isacharcolon}\ {\isachardoublequoteopen}{\isasymforall}R{\isachardot}\ trans\ R\ {\isasymand}\ eucl\ R\ {\isasymlongrightarrow}\ {\isacharparenleft}ser\ R\ {\isasymand}\ trans\ R\ {\isasymand}\ eucl\ R{\isacharparenright}{\isachardoublequoteclose}\ \isanewline
\ \isacommand{nitpick}\isamarkupfalse%
\isadelimproof
\ %
\endisadelimproof
\isatagproof
\isacommand{oops}\isamarkupfalse%
\endisatagproof
{\isafoldproof}%
\isadelimproof
\endisadelimproof
\isanewline
\isacommand{theorem}\isamarkupfalse%
\ C{\isadigit{1}}{\isadigit{1}}{\isacharunderscore}B{\isacharcolon}\ {\isachardoublequoteopen}{\isacharhash}\isactrlsup {\isadigit{1}}\ i{\isadigit{1}}\ {\isasymlongrightarrow}\ {\isasymnot}\ {\isacharparenleft}{\isasymforall}R{\isachardot}\ trans\ R\ {\isasymand}\ eucl\ R\ {\isasymlongrightarrow}\ {\isacharparenleft}ser\ R\ {\isasymand}\ trans\ R\ {\isasymand}\ eucl\ R{\isacharparenright}{\isacharparenright}{\isachardoublequoteclose}\isanewline
\isadelimproof
\ %
\endisadelimproof
\isatagproof
\isacommand{using}\isamarkupfalse%
\ B{\isadigit{9}}\ C{\isadigit{3}}{\isacharunderscore}D\ C{\isadigit{9}}{\isacharunderscore}B\ \isacommand{by}\isamarkupfalse%
\ blast%
\endisatagproof
{\isafoldproof}%
\isadelimproof
\endisadelimproof
\begin{isamarkuptext}%
\begin{isbfig}{7em}
\begin{tikzpicture}[shorten >=1pt,node distance=2cm,on grid,auto] 
   \node[state] (i_1)   {$i_1$}; 
\end{tikzpicture}
\end{isbfig}%
\end{isamarkuptext}%
\isamarkuptrue%
\isamarkupsubsubsection{DB $>$ KB%
}
\isamarkuptrue%
\isacommand{lemma}\isamarkupfalse%
\ C{\isadigit{1}}{\isadigit{2}}{\isacharunderscore}A{\isacharcolon}\ {\isachardoublequoteopen}{\isasymforall}R{\isachardot}\ sym\ R\ {\isasymlongrightarrow}\ {\isacharparenleft}ser\ R\ {\isasymand}\ sym\ R{\isacharparenright}{\isachardoublequoteclose}\ \isanewline
\ \isacommand{nitpick}\isamarkupfalse%
\isadelimproof
\ %
\endisadelimproof
\isatagproof
\isacommand{oops}\isamarkupfalse%
\endisatagproof
{\isafoldproof}%
\isadelimproof
\endisadelimproof
\isanewline
\isacommand{theorem}\isamarkupfalse%
\ C{\isadigit{1}}{\isadigit{2}}{\isacharunderscore}B{\isacharcolon}\ {\isachardoublequoteopen}{\isacharhash}\isactrlsup {\isadigit{1}}\ i{\isadigit{1}}\ {\isasymlongrightarrow}\ {\isasymnot}\ {\isacharparenleft}{\isasymforall}R{\isachardot}\ sym\ R\ {\isasymlongrightarrow}\ {\isacharparenleft}ser\ R\ {\isasymand}\ sym\ R{\isacharparenright}{\isacharparenright}{\isachardoublequoteclose}\ \isanewline
\isadelimproof
\ %
\endisadelimproof
\isatagproof
\isacommand{using}\isamarkupfalse%
\ C{\isadigit{1}}{\isadigit{1}}{\isacharunderscore}B\ C{\isadigit{3}}{\isacharunderscore}D\ \isacommand{by}\isamarkupfalse%
\ blast%
\endisatagproof
{\isafoldproof}%
\isadelimproof
\endisadelimproof
\begin{isamarkuptext}%
\begin{isbfig}{9em}
\begin{tikzpicture}[shorten >=1pt,node distance=2cm,on grid,auto] 
   \node[state] (i_1)   {$i_1$}; 
\end{tikzpicture}
\end{isbfig}%
\end{isamarkuptext}%
\isamarkuptrue%
\isamarkupsubsubsection{S5 $>$ KB5%
}
\isamarkuptrue%
\isacommand{lemma}\isamarkupfalse%
\ C{\isadigit{1}}{\isadigit{3}}{\isacharunderscore}A{\isacharcolon}\ {\isachardoublequoteopen}{\isasymforall}R{\isachardot}\ sym\ R\ {\isasymand}\ eucl\ R\ {\isasymlongrightarrow}\ {\isacharparenleft}refl\ R\ {\isasymand}\ eucl\ R{\isacharparenright}{\isachardoublequoteclose}\ \isanewline
\ \isacommand{nitpick}\isamarkupfalse%
\isadelimproof
\ %
\endisadelimproof
\isatagproof
\isacommand{oops}\isamarkupfalse%
\endisatagproof
{\isafoldproof}%
\isadelimproof
\endisadelimproof
\isanewline
\isacommand{theorem}\isamarkupfalse%
\ C{\isadigit{1}}{\isadigit{3}}{\isacharunderscore}B{\isacharcolon}\ {\isachardoublequoteopen}{\isacharhash}\isactrlsup {\isadigit{1}}\ i{\isadigit{1}}\ {\isasymlongrightarrow}\ {\isasymnot}\ {\isacharparenleft}{\isasymforall}R{\isachardot}\ sym\ R\ {\isasymand}\ eucl\ R\ {\isasymlongrightarrow}\ {\isacharparenleft}refl\ R\ {\isasymand}\ eucl\ R{\isacharparenright}{\isacharparenright}{\isachardoublequoteclose}%
\isadelimproof
\ %
\endisadelimproof
\isatagproof
\isacommand{using}\isamarkupfalse%
\ B{\isadigit{5}}\ C{\isadigit{1}}{\isadigit{2}}{\isacharunderscore}B\ C{\isadigit{6}}{\isacharunderscore}D\ \isacommand{by}\isamarkupfalse%
\ blast%
\endisatagproof
{\isafoldproof}%
\isadelimproof
\endisadelimproof
\begin{isamarkuptext}%
\begin{isbfig}{7em}
\begin{tikzpicture}[shorten >=1pt,node distance=2cm,on grid,auto] 
   \node[state] (i_1)   {$i_1$}; 
   \node[state] (i_2) [right=of i_1] {$i_2$}; 
    \path[->] 
    (i_1) edge [loop above] node {} ()
          edge [bend left] node {} (i_2)
    (i_2) edge [bend left] node {} (i_1);
\end{tikzpicture}
\end{isbfig}%
\end{isamarkuptext}%
\isamarkuptrue%
\isamarkupsubsubsection{D4 $>$ D%
}
\isamarkuptrue%
\isacommand{lemma}\isamarkupfalse%
\ C{\isadigit{1}}{\isadigit{4}}{\isacharunderscore}A{\isacharcolon}\ {\isachardoublequoteopen}{\isasymforall}R{\isachardot}\ {\isacharparenleft}ser\ R{\isacharparenright}\ {\isasymlongrightarrow}\ {\isacharparenleft}ser\ R{\isacharparenright}\ {\isasymand}\ {\isacharparenleft}trans\ R{\isacharparenright}{\isachardoublequoteclose}\ \isanewline
\ \isacommand{nitpick}\isamarkupfalse%
\isadelimproof
\ %
\endisadelimproof
\isatagproof
\isacommand{oops}\isamarkupfalse%
\endisatagproof
{\isafoldproof}%
\isadelimproof
\endisadelimproof
\isanewline
\isacommand{theorem}\isamarkupfalse%
\ C{\isadigit{1}}{\isadigit{4}}{\isacharunderscore}B{\isacharunderscore}Isabelle{\isacharunderscore}challenge{\isacharcolon}\ {\isachardoublequoteopen}{\isacharhash}\isactrlsup {\isadigit{2}}\ i{\isadigit{1}}\ i{\isadigit{2}}\ {\isasymlongrightarrow}\ {\isasymnot}{\isacharparenleft}{\isasymforall}R{\isachardot}\ ser\ R\ {\isasymlongrightarrow}\ {\isacharparenleft}ser\ R\ {\isasymand}\ trans\ R{\isacharparenright}{\isacharparenright}{\isachardoublequoteclose}\isanewline
\isamarkupcmt{sledgehammer [remote\_leo2] (ser\_def trans\_def) -- CPU time: 13.08 s. Metis reconstruction failed.%
}
\isanewline
\isamarkupcmt{sledgehammer [cvc4,timeout=300] suggested following line:%
}
\isanewline
\isadelimproof
\ %
\endisadelimproof
\isatagproof
\isacommand{by}\isamarkupfalse%
\ {\isacharparenleft}metis\ {\isacharparenleft}full{\isacharunderscore}types{\isacharparenright}\ C{\isadigit{1}}{\isacharunderscore}B\ trans{\isacharunderscore}def\ ser{\isacharunderscore}def{\isacharparenright}%
\endisatagproof
{\isafoldproof}%
\isadelimproof
\endisadelimproof
\ \isanewline
\isacommand{lemma}\isamarkupfalse%
\ C{\isadigit{1}}{\isadigit{4}}{\isacharunderscore}D{\isacharcolon}\ {\isachardoublequoteopen}{\isacharhash}\isactrlsup {\isadigit{1}}\ i{\isadigit{1}}\ {\isasymlongrightarrow}\ {\isacharparenleft}{\isasymforall}R{\isachardot}\ ser\ R\ {\isasymlongrightarrow}\ {\isacharparenleft}ser\ R\ {\isasymand}\ trans\ R{\isacharparenright}{\isacharparenright}{\isachardoublequoteclose}%
\isadelimproof
\ %
\endisadelimproof
\isatagproof
\isacommand{by}\isamarkupfalse%
\ {\isacharparenleft}metis\ {\isacharparenleft}full{\isacharunderscore}types{\isacharparenright}\ trans{\isacharunderscore}def{\isacharparenright}%
\endisatagproof
{\isafoldproof}%
\isadelimproof
\endisadelimproof
\isamarkupsubsubsection{D5 $>$ D%
}
\isamarkuptrue%
\begin{isamarkuptext}%
\begin{isbfig}{7em}
\begin{tikzpicture}[shorten >=1pt,node distance=2cm,on grid,auto] 
   \node[state] (i_1)   {$i_1$}; 
   \node[state] (i_2) [right=of i_1] {$i_2$}; 
    \path[->] 
    (i_1) edge [loop above] node {} ()
    (i_2) edge [loop above] node {} () 
          edge node {} (i_1);
\end{tikzpicture}
\end{isbfig}%
\end{isamarkuptext}%
\isamarkuptrue%
\isacommand{lemma}\isamarkupfalse%
\ C{\isadigit{1}}{\isadigit{5}}{\isacharunderscore}A{\isacharcolon}\ {\isachardoublequoteopen}{\isasymforall}R{\isachardot}\ ser\ R\ {\isasymlongrightarrow}\ {\isacharparenleft}ser\ R\ {\isasymand}\ eucl\ R{\isacharparenright}{\isachardoublequoteclose}\ \isanewline
\ \isacommand{nitpick}\isamarkupfalse%
\isadelimproof
\ %
\endisadelimproof
\isatagproof
\isacommand{oops}\isamarkupfalse%
\endisatagproof
{\isafoldproof}%
\isadelimproof
\endisadelimproof
\isanewline
\isacommand{theorem}\isamarkupfalse%
\ C{\isadigit{1}}{\isadigit{5}}{\isacharunderscore}B{\isacharunderscore}Isabelle{\isacharunderscore}challenge{\isacharcolon}\ {\isachardoublequoteopen}{\isacharhash}\isactrlsup {\isadigit{2}}\ i{\isadigit{1}}\ i{\isadigit{2}}\ {\isasymlongrightarrow}\ {\isasymnot}\ {\isacharparenleft}{\isasymforall}R{\isachardot}\ ser\ R\ {\isasymlongrightarrow}\ {\isacharparenleft}ser\ R\ {\isasymand}\ eucl\ R{\isacharparenright}{\isacharparenright}{\isachardoublequoteclose}\isanewline
\isamarkupcmt{sledgehammer [remote\_leo2](ser\_def eucl\_def)%
}
\isanewline
\isamarkupcmt{CPU time: 12.9 s. Metis reconstruction failed.%
}
\isanewline
\isamarkupcmt{sledgehammer [cvc4,timeout=300] suggested following line:%
}
\isanewline
\isadelimproof
\ %
\endisadelimproof
\isatagproof
\isacommand{by}\isamarkupfalse%
\ {\isacharparenleft}metis\ {\isacharparenleft}full{\isacharunderscore}types{\isacharparenright}\ C{\isadigit{1}}{\isadigit{4}}{\isacharunderscore}B{\isacharunderscore}Isabelle{\isacharunderscore}challenge\ trans{\isacharunderscore}def\ eucl{\isacharunderscore}def{\isacharparenright}%
\endisatagproof
{\isafoldproof}%
\isadelimproof
\endisadelimproof
\ \isanewline
\isacommand{lemma}\isamarkupfalse%
\ C{\isadigit{1}}{\isadigit{5}}{\isacharunderscore}D{\isacharcolon}\ {\isachardoublequoteopen}{\isacharhash}\isactrlsup {\isadigit{1}}\ i{\isadigit{1}}\ {\isasymlongrightarrow}\ {\isacharparenleft}{\isasymforall}R{\isachardot}\ ser\ R\ {\isasymlongrightarrow}\ {\isacharparenleft}ser\ R\ {\isasymand}\ eucl\ R{\isacharparenright}{\isacharparenright}{\isachardoublequoteclose}%
\isadelimproof
\ %
\endisadelimproof
\isatagproof
\isacommand{by}\isamarkupfalse%
\ {\isacharparenleft}metis\ {\isacharparenleft}full{\isacharunderscore}types{\isacharparenright}\ C{\isadigit{2}}{\isacharunderscore}D{\isacharparenright}%
\endisatagproof
{\isafoldproof}%
\isadelimproof
\endisadelimproof
\begin{isamarkuptext}%
\begin{isbfig}{7em}
\begin{tikzpicture}[shorten >=1pt,node distance=2cm,on grid,auto] 
   \node[state] (i_1)   {$i_1$}; 
   \node[state] (i_2) [right=of i_1] {$i_2$}; 
    \path[->] 
    (i_1) edge [loop above] node {} () 
          edge node {} (i_2)
    (i_2) edge [loop above] node {} ();
\end{tikzpicture}
\end{isbfig}%
\end{isamarkuptext}%
\isamarkuptrue%
\isamarkupsubsubsection{DB $>$ D%
}
\isamarkuptrue%
\isacommand{lemma}\isamarkupfalse%
\ C{\isadigit{1}}{\isadigit{6}}{\isacharunderscore}A{\isacharcolon}\ {\isachardoublequoteopen}{\isasymforall}R{\isachardot}\ ser\ R\ {\isasymlongrightarrow}\ {\isacharparenleft}ser\ R\ {\isasymand}\ sym\ R{\isacharparenright}{\isachardoublequoteclose}\ \isanewline
\ \isacommand{nitpick}\isamarkupfalse%
\isadelimproof
\ %
\endisadelimproof
\isatagproof
\isacommand{oops}\isamarkupfalse%
\endisatagproof
{\isafoldproof}%
\isadelimproof
\endisadelimproof
\isanewline
\isanewline
\isacommand{lemma}\isamarkupfalse%
\ C{\isadigit{1}}{\isadigit{6}}{\isacharunderscore}B{\isacharcolon}\ {\isachardoublequoteopen}{\isacharhash}\isactrlsup {\isadigit{2}}\ i{\isadigit{1}}\ i{\isadigit{2}}\ {\isasymlongrightarrow}\ {\isasymnot}\ {\isacharparenleft}{\isasymforall}R{\isachardot}\ ser\ R\ {\isasymlongrightarrow}\ {\isacharparenleft}ser\ R\ {\isasymand}\ sym\ R{\isacharparenright}{\isacharparenright}{\isachardoublequoteclose}%
\isadelimproof
\ %
\endisadelimproof
\isatagproof
\isacommand{by}\isamarkupfalse%
\ {\isacharparenleft}simp\ add{\isacharcolon}ser{\isacharunderscore}def\ sym{\isacharunderscore}def{\isacharcomma}\ force{\isacharparenright}%
\endisatagproof
{\isafoldproof}%
\isadelimproof
\endisadelimproof
\ \isanewline
\isacommand{lemma}\isamarkupfalse%
\ C{\isadigit{1}}{\isadigit{6}}{\isacharunderscore}D{\isacharcolon}\ {\isachardoublequoteopen}{\isacharhash}\isactrlsup {\isadigit{1}}\ i{\isadigit{1}}\ {\isasymlongrightarrow}\ {\isacharparenleft}{\isasymforall}R{\isachardot}\ ser\ R\ {\isasymlongrightarrow}\ {\isacharparenleft}ser\ R\ {\isasymand}\ sym\ R{\isacharparenright}{\isacharparenright}{\isachardoublequoteclose}%
\isadelimproof
\ %
\endisadelimproof
\isatagproof
\isacommand{by}\isamarkupfalse%
\ {\isacharparenleft}metis\ {\isacharparenleft}full{\isacharunderscore}types{\isacharparenright}\ sym{\isacharunderscore}def{\isacharparenright}%
\endisatagproof
{\isafoldproof}%
\isadelimproof
\endisadelimproof
\begin{isamarkuptext}%
\begin{isbfig}{7em}
\begin{tikzpicture}[shorten >=1pt,node distance=2cm,on grid,auto] 
   \node[state] (i_1)   {$i_1$}; 
   \node[state] (i_2) [right=of i_1] {$i_2$}; 
    \path[->] 
    (i_1) edge [loop above] node {} ()
          edge node {} (i_2)
    (i_2) edge [loop above] node {} ();
\end{tikzpicture}
\end{isbfig}%
\end{isamarkuptext}%
\isamarkuptrue%
\isamarkupsubsubsection{D45 $>$ D4%
}
\isamarkuptrue%
\isacommand{lemma}\isamarkupfalse%
\ C{\isadigit{1}}{\isadigit{7}}{\isacharunderscore}A{\isacharcolon}\ {\isachardoublequoteopen}{\isasymforall}R{\isachardot}\ ser\ R\ {\isasymand}\ trans\ R\ {\isasymlongrightarrow}\ {\isacharparenleft}ser\ R\ {\isasymand}\ trans\ R\ {\isasymand}\ eucl\ R{\isacharparenright}{\isachardoublequoteclose}\isanewline
\ \isacommand{nitpick}\isamarkupfalse%
\isadelimproof
\ %
\endisadelimproof
\isatagproof
\isacommand{oops}\isamarkupfalse%
\endisatagproof
{\isafoldproof}%
\isadelimproof
\endisadelimproof
\isanewline
\isanewline
\isacommand{lemma}\isamarkupfalse%
\ C{\isadigit{1}}{\isadigit{7}}{\isacharunderscore}B{\isacharunderscore}ATP{\isacharunderscore}challenge{\isacharcolon}\ {\isachardoublequoteopen}{\isacharhash}\isactrlsup {\isadigit{2}}\ i{\isadigit{1}}\ i{\isadigit{2}}\ {\isasymlongrightarrow}\ {\isasymnot}{\isacharparenleft}{\isasymforall}R{\isachardot}\ ser\ R\ {\isasymand}\ trans\ R\ {\isasymlongrightarrow}\ {\isacharparenleft}ser\ R\ {\isasymand}\ trans\ R\ {\isasymand}\ eucl\ R{\isacharparenright}{\isacharparenright}{\isachardoublequoteclose}\isanewline
\isadelimproof
\ %
\endisadelimproof
\isatagproof
\isacommand{oops}\isamarkupfalse%
\ %
\isamarkupcmt{All ATPs time out%
}
\endisatagproof
{\isafoldproof}%
\isadelimproof
\isanewline
\endisadelimproof
\isacommand{theorem}\isamarkupfalse%
\ C{\isadigit{1}}{\isadigit{7}}{\isacharunderscore}C{\isacharcolon}\ {\isachardoublequoteopen}{\isacharhash}\isactrlsup {\isadigit{2}}\ i{\isadigit{1}}\ i{\isadigit{2}}\ {\isasymand}\ r\ i{\isadigit{1}}\ i{\isadigit{1}}\ {\isasymand}\ r\ i{\isadigit{1}}\ i{\isadigit{2}}\ {\isasymand}\ {\isasymnot}\ r\ i{\isadigit{2}}\ i{\isadigit{1}}\ {\isasymand}\ r\ i{\isadigit{2}}\ i{\isadigit{2}}\ {\isasymlongrightarrow}\ {\isasymnot}\ {\isacharparenleft}ser\ r\ {\isasymand}\ trans\ r\ {\isasymlongrightarrow}\ {\isacharparenleft}ser\ r\ {\isasymand}\ trans\ r\ {\isasymand}\ eucl\ r{\isacharparenright}{\isacharparenright}{\isachardoublequoteclose}\ \isanewline
\isadelimproof
\ %
\endisadelimproof
\isatagproof
\isacommand{by}\isamarkupfalse%
\ {\isacharparenleft}metis\ {\isacharparenleft}full{\isacharunderscore}types{\isacharparenright}\ ser{\isacharunderscore}def\ trans{\isacharunderscore}def\ eucl{\isacharunderscore}def{\isacharparenright}%
\endisatagproof
{\isafoldproof}%
\isadelimproof
\isanewline
\endisadelimproof
\isacommand{lemma}\isamarkupfalse%
\ C{\isadigit{1}}{\isadigit{7}}{\isacharunderscore}D{\isacharcolon}\ {\isachardoublequoteopen}{\isacharhash}\isactrlsup {\isadigit{1}}\ i{\isadigit{1}}\ {\isasymlongrightarrow}\ {\isacharparenleft}{\isasymforall}R{\isachardot}\ ser\ R\ {\isasymand}\ trans\ R\ {\isasymlongrightarrow}\ {\isacharparenleft}ser\ R\ {\isasymand}\ trans\ R\ {\isasymand}\ eucl\ R{\isacharparenright}{\isacharparenright}{\isachardoublequoteclose}\isanewline
\isadelimproof
\ %
\endisadelimproof
\isatagproof
\isacommand{by}\isamarkupfalse%
\ {\isacharparenleft}metis\ {\isacharparenleft}full{\isacharunderscore}types{\isacharparenright}\ eucl{\isacharunderscore}def{\isacharparenright}%
\endisatagproof
{\isafoldproof}%
\isadelimproof
\endisadelimproof
\begin{isamarkuptext}%
\begin{isbfig}{7em}
\begin{tikzpicture}[shorten >=1pt,node distance=2cm,on grid,auto] 
   \node[state] (i_1)   {$i_1$}; 
   \node[state] (i_2) [right=of i_1] {$i_2$}; 
   \node[state] (i_3) [right=of i_2] {$i_3$}; 
    \path[->] 
    (i_1) edge [loop above] node {} ()
          edge [bend left] node {} (i_2) 
    (i_2) edge [loop above] node {} () 
          edge [bend left] node {} (i_1) 
    (i_3) edge node {} (i_2);
\end{tikzpicture}
\end{isbfig}%
\end{isamarkuptext}%
\isamarkuptrue%
\isamarkupsubsubsection{D45 $>$ D5%
}
\isamarkuptrue%
\isacommand{lemma}\isamarkupfalse%
\ C{\isadigit{1}}{\isadigit{8}}{\isacharunderscore}A{\isacharcolon}\ {\isachardoublequoteopen}{\isasymforall}R{\isachardot}\ ser\ R\ {\isasymand}\ eucl\ R\ {\isasymlongrightarrow}\ {\isacharparenleft}ser\ R\ {\isasymand}\ trans\ R\ {\isasymand}\ eucl\ R{\isacharparenright}{\isachardoublequoteclose}\isanewline
\ \isacommand{nitpick}\isamarkupfalse%
\isadelimproof
\ %
\endisadelimproof
\isatagproof
\isacommand{oops}\isamarkupfalse%
\endisatagproof
{\isafoldproof}%
\isadelimproof
\endisadelimproof
\isanewline
\isanewline
\isacommand{lemma}\isamarkupfalse%
\ C{\isadigit{1}}{\isadigit{8}}{\isacharunderscore}ATP{\isacharunderscore}challenge{\isacharcolon}\ {\isachardoublequoteopen}{\isacharhash}\isactrlsup {\isadigit{3}}\ i{\isadigit{1}}\ i{\isadigit{2}}\ i{\isadigit{3}}\ {\isasymlongrightarrow}\ {\isasymnot}\ {\isacharparenleft}{\isasymforall}R{\isachardot}\ ser\ R\ {\isasymand}\ eucl\ R\ {\isasymlongrightarrow}\ {\isacharparenleft}ser\ R\ {\isasymand}\ trans\ R\ {\isasymand}\ eucl\ R{\isacharparenright}{\isacharparenright}{\isachardoublequoteclose}\isanewline
\isadelimproof
\ %
\endisadelimproof
\isatagproof
\isacommand{oops}\isamarkupfalse%
\ %
\isamarkupcmt{All ATPs time out%
}
\endisatagproof
{\isafoldproof}%
\isadelimproof
\isanewline
\endisadelimproof
\isacommand{theorem}\isamarkupfalse%
\ C{\isadigit{1}}{\isadigit{8}}{\isacharunderscore}C{\isacharcolon}\ {\isachardoublequoteopen}{\isacharhash}\isactrlsup {\isadigit{3}}\ i{\isadigit{1}}\ i{\isadigit{2}}\ i{\isadigit{3}}\ {\isasymand}\ r\ i{\isadigit{1}}\ i{\isadigit{1}}\ {\isasymand}\ r\ i{\isadigit{1}}\ i{\isadigit{2}}\ {\isasymand}\ {\isasymnot}\ r\ i{\isadigit{1}}\ i{\isadigit{3}}\ {\isasymand}\ r\ i{\isadigit{2}}\ i{\isadigit{1}}\ {\isasymand}\ r\ i{\isadigit{2}}\ i{\isadigit{2}}\ {\isasymand}\ {\isasymnot}\ r\ i{\isadigit{2}}\ i{\isadigit{3}}\ {\isasymand}\ {\isasymnot}\ r\ i{\isadigit{3}}\ i{\isadigit{1}}\ {\isasymand}\ r\ i{\isadigit{3}}\ i{\isadigit{2}}\ {\isasymand}\ {\isasymnot}\ r\ i{\isadigit{3}}\ i{\isadigit{3}}\ \ {\isasymlongrightarrow}\ {\isasymnot}\ {\isacharparenleft}ser\ r\ {\isasymand}\ eucl\ r\ {\isasymlongrightarrow}\ {\isacharparenleft}ser\ r\ {\isasymand}\ trans\ r\ {\isasymand}\ eucl\ r{\isacharparenright}{\isacharparenright}{\isachardoublequoteclose}%
\isadelimproof
\ %
\endisadelimproof
\isatagproof
\isacommand{by}\isamarkupfalse%
\ {\isacharparenleft}metis\ {\isacharparenleft}full{\isacharunderscore}types{\isacharparenright}\ eucl{\isacharunderscore}def\ ser{\isacharunderscore}def\ trans{\isacharunderscore}def{\isacharparenright}%
\endisatagproof
{\isafoldproof}%
\isadelimproof
\endisadelimproof
\isanewline
\isacommand{lemma}\isamarkupfalse%
\ C{\isadigit{1}}{\isadigit{8}}{\isacharunderscore}D{\isacharcolon}\ {\isachardoublequoteopen}{\isacharhash}\isactrlsup {\isadigit{2}}\ i{\isadigit{1}}\ i{\isadigit{2}}\ {\isasymlongrightarrow}\ {\isacharparenleft}{\isasymforall}R{\isachardot}\ ser\ R\ {\isasymand}\ eucl\ R\ {\isasymlongrightarrow}\ {\isacharparenleft}ser\ R\ {\isasymand}\ trans\ R\ {\isasymand}\ eucl\ R{\isacharparenright}{\isacharparenright}{\isachardoublequoteclose}\isanewline
\isadelimproof
\ %
\endisadelimproof
\isatagproof
\isacommand{by}\isamarkupfalse%
\ {\isacharparenleft}metis\ {\isacharparenleft}full{\isacharunderscore}types{\isacharparenright}\ eucl{\isacharunderscore}def\ trans{\isacharunderscore}def{\isacharparenright}%
\endisatagproof
{\isafoldproof}%
\isadelimproof
\endisadelimproof
\begin{isamarkuptext}%
\begin{isbfig}{7em}
\begin{tikzpicture}[shorten >=1pt,node distance=2cm,on grid,auto] 
   \node[state] (i_1)   {$i_1$}; 
   \node[state] (i_2) [right=of i_1] {$i_2$}; 
    \path[->] 
    (i_1) edge [loop above] node {} ()
    (i_2) edge node {} (i_1);
\end{tikzpicture}
\end{isbfig}%
\end{isamarkuptext}%
\isamarkuptrue%
\isamarkupsubsubsection{M $>$ D%
}
\isamarkuptrue%
\isacommand{lemma}\isamarkupfalse%
\ C{\isadigit{1}}{\isadigit{9}}{\isacharunderscore}A{\isacharcolon}\ {\isachardoublequoteopen}{\isasymforall}R{\isachardot}\ ser\ R\ {\isasymlongrightarrow}\ refl\ R{\isachardoublequoteclose}\ \isanewline
\ \isacommand{nitpick}\isamarkupfalse%
\isadelimproof
\ %
\endisadelimproof
\isatagproof
\isacommand{oops}\isamarkupfalse%
\endisatagproof
{\isafoldproof}%
\isadelimproof
\endisadelimproof
\isanewline
\isacommand{theorem}\isamarkupfalse%
\ C{\isadigit{1}}{\isadigit{9}}{\isacharunderscore}B{\isacharunderscore}Isabelle{\isacharunderscore}challenge{\isacharcolon}\ {\isachardoublequoteopen}{\isacharhash}\isactrlsup {\isadigit{2}}\ i{\isadigit{1}}\ i{\isadigit{2}}\ {\isasymlongrightarrow}\ {\isasymnot}\ {\isacharparenleft}{\isasymforall}R{\isachardot}\ ser\ R\ {\isasymlongrightarrow}\ refl\ R{\isacharparenright}{\isachardoublequoteclose}\isanewline
\isamarkupcmt{sledgehammer [remote\_leo2,timeout=200] (ser\_def refl\_def) -- CPU time: 29.15 s. Metis reconstruction failed.%
}
\isanewline
\isamarkupcmt{sledgehammer [cvc4,timeout=300] suggested following line:%
}
\isanewline
\isadelimproof
\ %
\endisadelimproof
\isatagproof
\isacommand{by}\isamarkupfalse%
\ {\isacharparenleft}metis\ {\isacharparenleft}full{\isacharunderscore}types{\isacharparenright}\ C{\isadigit{1}}{\isadigit{4}}{\isacharunderscore}B{\isacharunderscore}Isabelle{\isacharunderscore}challenge\ trans{\isacharunderscore}def\ refl{\isacharunderscore}def{\isacharparenright}%
\endisatagproof
{\isafoldproof}%
\isadelimproof
\endisadelimproof
\ \isanewline
\isacommand{lemma}\isamarkupfalse%
\ C{\isadigit{1}}{\isadigit{9}}{\isacharunderscore}D{\isacharcolon}\ {\isachardoublequoteopen}{\isacharhash}\isactrlsup {\isadigit{1}}\ i{\isadigit{1}}\ {\isasymlongrightarrow}\ {\isacharparenleft}{\isasymforall}R{\isachardot}\ ser\ R\ {\isasymlongrightarrow}\ refl\ R{\isacharparenright}{\isachardoublequoteclose}%
\isadelimproof
\ %
\endisadelimproof
\isatagproof
\isacommand{by}\isamarkupfalse%
\ {\isacharparenleft}metis\ {\isacharparenleft}full{\isacharunderscore}types{\isacharparenright}\ ser{\isacharunderscore}def\ refl{\isacharunderscore}def{\isacharparenright}%
\endisatagproof
{\isafoldproof}%
\isadelimproof
\endisadelimproof
\begin{isamarkuptext}%
\begin{isbfig}{7em}
\begin{tikzpicture}[shorten >=1pt,node distance=2cm,on grid,auto] 
   \node[state] (i_1)   {$i_1$}; 
   \node[state] (i_2) [right=of i_1] {$i_2$}; 
    \path[->] 
    (i_1) edge [loop above] node {} ()
    (i_2) edge node {} (i_1);
\end{tikzpicture}
\end{isbfig}%
\end{isamarkuptext}%
\isamarkuptrue%
\isamarkupsubsubsection{S4 $>$ D4%
}
\isamarkuptrue%
\isacommand{lemma}\isamarkupfalse%
\ C{\isadigit{2}}{\isadigit{0}}{\isacharunderscore}A{\isacharcolon}\ {\isachardoublequoteopen}{\isasymforall}R{\isachardot}\ ser\ R\ {\isasymand}\ trans\ R\ {\isasymlongrightarrow}\ {\isacharparenleft}refl\ R\ {\isasymand}\ trans\ R{\isacharparenright}{\isachardoublequoteclose}\isanewline
\ \isacommand{nitpick}\isamarkupfalse%
\isadelimproof
\ %
\endisadelimproof
\isatagproof
\isacommand{oops}\isamarkupfalse%
\endisatagproof
{\isafoldproof}%
\isadelimproof
\endisadelimproof
\isanewline
\isanewline
\isacommand{lemma}\isamarkupfalse%
\ C{\isadigit{2}}{\isadigit{0}}{\isacharunderscore}B{\isacharunderscore}Isabelle{\isacharunderscore}challenge{\isacharcolon}\ {\isachardoublequoteopen}{\isacharhash}\isactrlsup {\isadigit{2}}\ i{\isadigit{1}}\ i{\isadigit{2}}\ {\isasymlongrightarrow}\ {\isasymnot}\ {\isacharparenleft}{\isasymforall}R{\isachardot}\ ser\ R\ {\isasymand}\ trans\ R\ {\isasymlongrightarrow}\ {\isacharparenleft}refl\ R\ {\isasymand}\ trans\ R{\isacharparenright}{\isacharparenright}{\isachardoublequoteclose}\isanewline
\isamarkupcmt{sledgehammer [remote\_leo2](ser\_def trans\_def refl\_def) -- CPU time: 12.5 s. Metis reconstruction failed.%
}
\isanewline
\isamarkupcmt{sledgehammer [cvc4,timeout=300] -- timed out%
}
\isanewline
\isadelimproof
\ %
\endisadelimproof
\isatagproof
\isacommand{oops}\isamarkupfalse%
\endisatagproof
{\isafoldproof}%
\isadelimproof
\isanewline
\endisadelimproof
\isacommand{theorem}\isamarkupfalse%
\ C{\isadigit{2}}{\isadigit{0}}{\isacharunderscore}C{\isacharcolon}\ {\isachardoublequoteopen}{\isacharhash}\isactrlsup {\isadigit{2}}\ i{\isadigit{1}}\ i{\isadigit{2}}\ {\isasymand}\ r\ i{\isadigit{1}}\ i{\isadigit{1}}\ {\isasymand}\ {\isasymnot}\ r\ i{\isadigit{1}}\ i{\isadigit{2}}\ {\isasymand}\ r\ i{\isadigit{2}}\ i{\isadigit{1}}\ {\isasymand}\ {\isasymnot}\ r\ i{\isadigit{2}}\ i{\isadigit{2}}\ {\isasymlongrightarrow}\ {\isasymnot}\ {\isacharparenleft}ser\ r\ {\isasymand}\ trans\ r\ {\isasymlongrightarrow}\ {\isacharparenleft}refl\ r\ {\isasymand}\ trans\ r{\isacharparenright}{\isacharparenright}{\isachardoublequoteclose}\ \isanewline
\isadelimproof
\ %
\endisadelimproof
\isatagproof
\isacommand{by}\isamarkupfalse%
\ {\isacharparenleft}metis\ {\isacharparenleft}full{\isacharunderscore}types{\isacharparenright}\ ser{\isacharunderscore}def\ refl{\isacharunderscore}def\ trans{\isacharunderscore}def{\isacharparenright}%
\endisatagproof
{\isafoldproof}%
\isadelimproof
\isanewline
\endisadelimproof
\isacommand{lemma}\isamarkupfalse%
\ C{\isadigit{2}}{\isadigit{0}}{\isacharunderscore}D{\isacharcolon}\ {\isachardoublequoteopen}{\isacharhash}\isactrlsup {\isadigit{1}}\ i{\isadigit{1}}\ \ {\isasymlongrightarrow}\ {\isacharparenleft}{\isasymforall}R{\isachardot}\ ser\ R\ {\isasymand}\ trans\ R\ {\isasymlongrightarrow}\ {\isacharparenleft}refl\ R\ {\isasymand}\ trans\ R{\isacharparenright}{\isacharparenright}{\isachardoublequoteclose}\isanewline
\isadelimproof
\ %
\endisadelimproof
\isatagproof
\isacommand{by}\isamarkupfalse%
\ {\isacharparenleft}metis\ {\isacharparenleft}full{\isacharunderscore}types{\isacharparenright}\ ser{\isacharunderscore}def\ refl{\isacharunderscore}def{\isacharparenright}%
\endisatagproof
{\isafoldproof}%
\isadelimproof
\endisadelimproof
\begin{isamarkuptext}%
\begin{isbfig}{7em}
\begin{tikzpicture}[shorten >=1pt,node distance=2cm,on grid,auto] 
   \node[state] (i_1)   {$i_1$}; 
   \node[state] (i_2) [right=of i_1] {$i_2$}; 
    \path[->] 
    (i_1) edge [loop above] node {} ()
    (i_2) edge node {} (i_1);
\end{tikzpicture}
\end{isbfig}%
\end{isamarkuptext}%
\isamarkuptrue%
\isamarkupsubsubsection{S5 $>$ D45%
}
\isamarkuptrue%
\isacommand{lemma}\isamarkupfalse%
\ C{\isadigit{2}}{\isadigit{1}}{\isacharunderscore}A{\isacharcolon}\ {\isachardoublequoteopen}{\isasymforall}R{\isachardot}\ ser\ R\ {\isasymand}\ trans\ R\ {\isasymand}\ eucl\ R\ {\isasymlongrightarrow}\ {\isacharparenleft}refl\ R\ {\isasymand}\ eucl\ R{\isacharparenright}{\isachardoublequoteclose}\ \isanewline
\ \isacommand{nitpick}\isamarkupfalse%
\isadelimproof
\ %
\endisadelimproof
\isatagproof
\isacommand{oops}\isamarkupfalse%
\endisatagproof
{\isafoldproof}%
\isadelimproof
\endisadelimproof
\isanewline
\isanewline
\isacommand{lemma}\isamarkupfalse%
\ C{\isadigit{2}}{\isadigit{1}}{\isacharunderscore}B{\isacharunderscore}Isabelle{\isacharunderscore}challenge{\isacharcolon}\ {\isachardoublequoteopen}{\isacharhash}\isactrlsup {\isadigit{2}}\ i{\isadigit{1}}\ i{\isadigit{2}}\ {\isasymlongrightarrow}\ {\isasymnot}\ {\isacharparenleft}{\isasymforall}R{\isachardot}\ ser\ R\ {\isasymand}\ trans\ R\ {\isasymand}\ eucl\ R\ {\isasymlongrightarrow}\ {\isacharparenleft}refl\ R\ {\isasymand}\ eucl\ R{\isacharparenright}{\isacharparenright}{\isachardoublequoteclose}\isanewline
\isamarkupcmt{sledgehammer [remote\_leo2](ser\_def trans\_def eucl\_def refl\_def) -- CPU time: 12.51 s. Metis reconstruction failed.%
}
\isanewline
\isamarkupcmt{sledgehammer [cvc4,timeout=300] -- timed out%
}
\isanewline
\isadelimproof
\ %
\endisadelimproof
\isatagproof
\isacommand{oops}\isamarkupfalse%
\endisatagproof
{\isafoldproof}%
\isadelimproof
\isanewline
\endisadelimproof
\isacommand{theorem}\isamarkupfalse%
\ C{\isadigit{2}}{\isadigit{1}}{\isacharunderscore}C{\isacharcolon}\ {\isachardoublequoteopen}{\isacharhash}\isactrlsup {\isadigit{2}}\ i{\isadigit{1}}\ i{\isadigit{2}}\ {\isasymand}\ r\ i{\isadigit{1}}\ i{\isadigit{1}}\ {\isasymand}\ {\isasymnot}\ r\ i{\isadigit{1}}\ i{\isadigit{2}}\ {\isasymand}\ r\ i{\isadigit{2}}\ i{\isadigit{1}}\ {\isasymand}\ {\isasymnot}\ r\ i{\isadigit{2}}\ i{\isadigit{2}}\ {\isasymlongrightarrow}\ {\isasymnot}\ {\isacharparenleft}ser\ r\ {\isasymand}\ trans\ r\ {\isasymand}\ eucl\ r\ {\isasymlongrightarrow}\ {\isacharparenleft}refl\ r\ {\isasymand}\ eucl\ r{\isacharparenright}{\isacharparenright}{\isachardoublequoteclose}\ \isanewline
\isadelimproof
\ %
\endisadelimproof
\isatagproof
\isacommand{by}\isamarkupfalse%
\ {\isacharparenleft}metis\ {\isacharparenleft}full{\isacharunderscore}types{\isacharparenright}\ ser{\isacharunderscore}def\ trans{\isacharunderscore}def\ eucl{\isacharunderscore}def\ refl{\isacharunderscore}def{\isacharparenright}%
\endisatagproof
{\isafoldproof}%
\isadelimproof
\isanewline
\endisadelimproof
\isacommand{lemma}\isamarkupfalse%
\ C{\isadigit{2}}{\isadigit{1}}{\isacharunderscore}inclusion{\isacharcolon}\ {\isachardoublequoteopen}{\isacharhash}\isactrlsup {\isadigit{1}}\ i{\isadigit{1}}\ {\isasymlongrightarrow}\ {\isacharparenleft}{\isasymforall}R{\isachardot}\ ser\ R\ {\isasymand}\ trans\ R\ {\isasymand}\ eucl\ R\ {\isasymlongrightarrow}\ {\isacharparenleft}refl\ R\ {\isasymand}\ eucl\ R{\isacharparenright}{\isacharparenright}{\isachardoublequoteclose}\isanewline
\isadelimproof
\ %
\endisadelimproof
\isatagproof
\isacommand{by}\isamarkupfalse%
\ {\isacharparenleft}metis\ {\isacharparenleft}full{\isacharunderscore}types{\isacharparenright}\ ser{\isacharunderscore}def\ refl{\isacharunderscore}def{\isacharparenright}%
\endisatagproof
{\isafoldproof}%
\isadelimproof
\endisadelimproof
\begin{isamarkuptext}%
\begin{isbfig}{7em}
\begin{tikzpicture}[shorten >=1pt,node distance=2cm,on grid,auto] 
   \node[state] (i_1)   {$i_1$}; 
   \node[state] (i_2) [right=of i_1] {$i_2$}; 
    \path[->] 
    (i_1) edge [loop above] node {} () 
          edge [bend left] node {} (i_2)
    (i_2) edge [bend left] node {} (i_1);
\end{tikzpicture}
\end{isbfig}%
\end{isamarkuptext}%
\isamarkuptrue%
\isamarkupsubsubsection{B $>$ DB%
}
\isamarkuptrue%
\isacommand{lemma}\isamarkupfalse%
\ C{\isadigit{2}}{\isadigit{2}}{\isacharunderscore}A{\isacharcolon}\ {\isachardoublequoteopen}{\isasymforall}R{\isachardot}\ ser\ R\ {\isasymand}\ sym\ R\ {\isasymlongrightarrow}\ {\isacharparenleft}refl\ R\ {\isasymand}\ sym\ R{\isacharparenright}{\isachardoublequoteclose}\ \isanewline
\ \isacommand{nitpick}\isamarkupfalse%
\isadelimproof
\ %
\endisadelimproof
\isatagproof
\isacommand{oops}\isamarkupfalse%
\endisatagproof
{\isafoldproof}%
\isadelimproof
\endisadelimproof
\isanewline
\isacommand{lemma}\isamarkupfalse%
\ C{\isadigit{2}}{\isadigit{2}}{\isacharunderscore}B{\isacharunderscore}Isabelle{\isacharunderscore}challenge{\isacharcolon}\ {\isachardoublequoteopen}{\isacharhash}\isactrlsup {\isadigit{2}}\ i{\isadigit{1}}\ i{\isadigit{2}}\ {\isasymlongrightarrow}\ {\isasymnot}\ {\isacharparenleft}{\isasymforall}R{\isachardot}\ ser\ R\ {\isasymand}\ sym\ R\ {\isasymlongrightarrow}\ {\isacharparenleft}refl\ R\ {\isasymand}\ sym\ R{\isacharparenright}{\isacharparenright}{\isachardoublequoteclose}\isanewline
\isamarkupcmt{sledgehammer [remote\_leo2,timeout=200](ser\_def sym\_def refl\_def) -- CPU time: 31.18 s. Metis reconstruction failed.%
}
\isanewline
\isamarkupcmt{sledgehammer [cvc4,timeout=300] suggested following line:%
}
\isanewline
\isamarkupcmt{by (smt C14\_B sym\_def trans\_def refl\_def)%
}
\isadelimproof
\ %
\endisadelimproof
\isatagproof
\isacommand{oops}\isamarkupfalse%
\endisatagproof
{\isafoldproof}%
\isadelimproof
\endisadelimproof
\isanewline
\isacommand{theorem}\isamarkupfalse%
\ C{\isadigit{2}}{\isadigit{2}}{\isacharunderscore}C{\isacharcolon}\ {\isachardoublequoteopen}{\isacharhash}\isactrlsup {\isadigit{2}}\ i{\isadigit{1}}\ i{\isadigit{2}}\ {\isasymand}\ r\ i{\isadigit{1}}\ i{\isadigit{1}}\ {\isasymand}\ r\ i{\isadigit{1}}\ i{\isadigit{2}}\ {\isasymand}\ r\ i{\isadigit{2}}\ i{\isadigit{1}}\ {\isasymand}\ {\isasymnot}\ r\ i{\isadigit{2}}\ i{\isadigit{2}}\ {\isasymlongrightarrow}\ {\isasymnot}\ {\isacharparenleft}ser\ r\ {\isasymand}\ sym\ r\ {\isasymlongrightarrow}\ {\isacharparenleft}refl\ r\ {\isasymand}\ sym\ r{\isacharparenright}{\isacharparenright}{\isachardoublequoteclose}\ \isanewline
\isadelimproof
\ %
\endisadelimproof
\isatagproof
\isacommand{by}\isamarkupfalse%
\ {\isacharparenleft}metis\ {\isacharparenleft}full{\isacharunderscore}types{\isacharparenright}\ ser{\isacharunderscore}def\ sym{\isacharunderscore}def\ refl{\isacharunderscore}def{\isacharparenright}%
\endisatagproof
{\isafoldproof}%
\isadelimproof
\isanewline
\endisadelimproof
\isacommand{lemma}\isamarkupfalse%
\ C{\isadigit{2}}{\isadigit{2}}{\isacharunderscore}D{\isacharcolon}\ {\isachardoublequoteopen}{\isacharhash}\isactrlsup {\isadigit{1}}\ i{\isadigit{1}}\ {\isasymlongrightarrow}\ {\isacharparenleft}{\isasymforall}R{\isachardot}\ ser\ R\ {\isasymand}\ sym\ R\ {\isasymlongrightarrow}\ {\isacharparenleft}refl\ R\ {\isasymand}\ sym\ R{\isacharparenright}{\isacharparenright}{\isachardoublequoteclose}\isanewline
\isadelimproof
\ %
\endisadelimproof
\isatagproof
\isacommand{by}\isamarkupfalse%
\ {\isacharparenleft}metis\ {\isacharparenleft}full{\isacharunderscore}types{\isacharparenright}\ ser{\isacharunderscore}def\ refl{\isacharunderscore}def{\isacharparenright}%
\endisatagproof
{\isafoldproof}%
\isadelimproof
\endisadelimproof
\begin{isamarkuptext}%
\begin{isbfig}{7em}
\begin{tikzpicture}[shorten >=1pt,node distance=2cm,on grid,auto] 
   \node[state] (i_1)   {$i_1$}; 
   \node[state] (i_2) [right=of i_1] {$i_2$}; 
    \path[->] 
    (i_1) edge [loop above] node {} () 
          edge node {} (i_2) 
    (i_2) edge [loop above] node {} ();
\end{tikzpicture}
\end{isbfig}%
\end{isamarkuptext}%
\isamarkuptrue%
\isamarkupsubsubsection{B $>$ M%
}
\isamarkuptrue%
\isacommand{lemma}\isamarkupfalse%
\ C{\isadigit{2}}{\isadigit{3}}{\isacharunderscore}A{\isacharcolon}\ {\isachardoublequoteopen}{\isasymforall}R{\isachardot}\ refl\ R\ {\isasymlongrightarrow}\ {\isacharparenleft}refl\ R\ {\isasymand}\ sym\ R{\isacharparenright}{\isachardoublequoteclose}\ \isacommand{nitpick}\isamarkupfalse%
\isadelimproof
\ %
\endisadelimproof
\isatagproof
\isacommand{oops}\isamarkupfalse%
\endisatagproof
{\isafoldproof}%
\isadelimproof
\endisadelimproof
\isanewline
\isacommand{lemma}\isamarkupfalse%
\ C{\isadigit{2}}{\isadigit{3}}{\isacharunderscore}B{\isacharunderscore}ATP{\isacharunderscore}challenge{\isacharcolon}\ {\isachardoublequoteopen}{\isacharhash}\isactrlsup {\isadigit{2}}\ i{\isadigit{1}}\ i{\isadigit{2}}\ {\isasymlongrightarrow}\ {\isasymnot}\ {\isacharparenleft}{\isasymforall}R{\isachardot}\ refl\ R\ {\isasymlongrightarrow}\ {\isacharparenleft}refl\ R\ {\isasymand}\ sym\ R{\isacharparenright}{\isacharparenright}{\isachardoublequoteclose}\isanewline
\isadelimproof
\ %
\endisadelimproof
\isatagproof
\isacommand{oops}\isamarkupfalse%
\ %
\isamarkupcmt{All ATPs time out%
}
\endisatagproof
{\isafoldproof}%
\isadelimproof
\isanewline
\endisadelimproof
\isacommand{theorem}\isamarkupfalse%
\ C{\isadigit{2}}{\isadigit{3}}{\isacharunderscore}C{\isacharcolon}\ {\isachardoublequoteopen}{\isacharhash}\isactrlsup {\isadigit{2}}\ i{\isadigit{1}}\ i{\isadigit{2}}\ \ {\isasymand}\ r\ i{\isadigit{1}}\ i{\isadigit{1}}\ {\isasymand}\ r\ i{\isadigit{1}}\ i{\isadigit{2}}\ {\isasymand}\ {\isasymnot}\ r\ i{\isadigit{2}}\ i{\isadigit{1}}\ {\isasymand}\ r\ i{\isadigit{2}}\ i{\isadigit{2}}\ \ {\isasymlongrightarrow}\ {\isasymnot}\ {\isacharparenleft}refl\ r\ {\isasymlongrightarrow}\ {\isacharparenleft}refl\ r\ {\isasymand}\ sym\ r{\isacharparenright}{\isacharparenright}{\isachardoublequoteclose}\ \isanewline
\isadelimproof
\ %
\endisadelimproof
\isatagproof
\isacommand{by}\isamarkupfalse%
\ {\isacharparenleft}metis\ refl{\isacharunderscore}def\ sym{\isacharunderscore}def{\isacharparenright}%
\endisatagproof
{\isafoldproof}%
\isadelimproof
\isanewline
\endisadelimproof
\isacommand{lemma}\isamarkupfalse%
\ C{\isadigit{2}}{\isadigit{3}}{\isacharunderscore}D{\isacharcolon}\ {\isachardoublequoteopen}{\isacharhash}\isactrlsup {\isadigit{1}}\ i{\isadigit{1}}\ {\isasymlongrightarrow}\ {\isacharparenleft}{\isasymforall}R{\isachardot}\ refl\ R\ {\isasymlongrightarrow}\ {\isacharparenleft}refl\ R\ {\isasymand}\ sym\ R{\isacharparenright}{\isacharparenright}{\isachardoublequoteclose}%
\isadelimproof
\ %
\endisadelimproof
\isatagproof
\isacommand{by}\isamarkupfalse%
\ {\isacharparenleft}metis\ {\isacharparenleft}full{\isacharunderscore}types{\isacharparenright}\ sym{\isacharunderscore}def{\isacharparenright}%
\endisatagproof
{\isafoldproof}%
\isadelimproof
\endisadelimproof
\begin{isamarkuptext}%
\begin{isbfig}{7em}
\begin{tikzpicture}[shorten >=1pt,node distance=2cm,on grid,auto] 
   \node[state] (i_1)   {$i_1$}; 
   \node[state] (i_2) [right=of i_1] {$i_2$}; 
    \path[->] 
    (i_1) edge [loop above] node {} ()
          edge node {} (i_2)
    (i_2) edge [loop above] node {} ();
\end{tikzpicture}
\end{isbfig}%
\end{isamarkuptext}%
\isamarkuptrue%
\isamarkupsubsubsection{S5 $>$ S4%
}
\isamarkuptrue%
\isacommand{lemma}\isamarkupfalse%
\ C{\isadigit{2}}{\isadigit{4}}{\isacharunderscore}A{\isacharcolon}\ {\isachardoublequoteopen}{\isasymforall}R{\isachardot}\ refl\ R\ {\isasymand}\ trans\ R\ {\isasymlongrightarrow}\ {\isacharparenleft}refl\ R\ {\isasymand}\ eucl\ R{\isacharparenright}{\isachardoublequoteclose}\isanewline
\ \isacommand{nitpick}\isamarkupfalse%
\isadelimproof
\ %
\endisadelimproof
\isatagproof
\isacommand{oops}\isamarkupfalse%
\endisatagproof
{\isafoldproof}%
\isadelimproof
\endisadelimproof
\isanewline
\isanewline
\isacommand{lemma}\isamarkupfalse%
\ C{\isadigit{2}}{\isadigit{4}}{\isacharunderscore}B{\isacharunderscore}ATP{\isacharunderscore}challenge{\isacharcolon}\ {\isachardoublequoteopen}{\isacharhash}\isactrlsup {\isadigit{2}}\ i{\isadigit{1}}\ i{\isadigit{2}}\ {\isasymlongrightarrow}\ {\isasymnot}\ {\isacharparenleft}{\isasymforall}R{\isachardot}\ refl\ R\ {\isasymand}\ trans\ R\ {\isasymlongrightarrow}\ {\isacharparenleft}refl\ R\ {\isasymand}\ eucl\ R{\isacharparenright}{\isacharparenright}{\isachardoublequoteclose}\isanewline
\isadelimproof
\ %
\endisadelimproof
\isatagproof
\isacommand{oops}\isamarkupfalse%
\ %
\isamarkupcmt{All ATPs time out%
}
\endisatagproof
{\isafoldproof}%
\isadelimproof
\isanewline
\endisadelimproof
\isacommand{theorem}\isamarkupfalse%
\ C{\isadigit{2}}{\isadigit{4}}{\isacharunderscore}C{\isacharcolon}\ {\isachardoublequoteopen}{\isacharhash}\isactrlsup {\isadigit{2}}\ i{\isadigit{1}}\ i{\isadigit{2}}\ {\isasymand}\ r\ i{\isadigit{1}}\ i{\isadigit{1}}\ {\isasymand}\ r\ i{\isadigit{1}}\ i{\isadigit{2}}\ {\isasymand}\ {\isasymnot}\ r\ i{\isadigit{2}}\ i{\isadigit{1}}\ {\isasymand}\ r\ i{\isadigit{2}}\ i{\isadigit{2}}\ {\isasymlongrightarrow}\ {\isasymnot}\ {\isacharparenleft}refl\ r\ {\isasymand}\ trans\ r\ {\isasymlongrightarrow}\ {\isacharparenleft}refl\ r\ {\isasymand}\ eucl\ r{\isacharparenright}{\isacharparenright}{\isachardoublequoteclose}\ \isanewline
\isadelimproof
\ %
\endisadelimproof
\isatagproof
\isacommand{by}\isamarkupfalse%
\ {\isacharparenleft}metis\ {\isacharparenleft}full{\isacharunderscore}types{\isacharparenright}\ trans{\isacharunderscore}def\ refl{\isacharunderscore}def\ eucl{\isacharunderscore}def{\isacharparenright}%
\endisatagproof
{\isafoldproof}%
\isadelimproof
\isanewline
\endisadelimproof
\isacommand{lemma}\isamarkupfalse%
\ C{\isadigit{2}}{\isadigit{4}}{\isacharunderscore}D{\isacharcolon}\ {\isachardoublequoteopen}{\isacharhash}\isactrlsup {\isadigit{1}}\ i{\isadigit{1}}\ {\isasymlongrightarrow}\ {\isacharparenleft}{\isasymforall}R{\isachardot}\ refl\ R\ {\isasymand}\ trans\ R\ {\isasymlongrightarrow}\ {\isacharparenleft}refl\ R\ {\isasymand}\ eucl\ R{\isacharparenright}{\isacharparenright}{\isachardoublequoteclose}%
\isadelimproof
\ %
\endisadelimproof
\isatagproof
\isacommand{by}\isamarkupfalse%
\ {\isacharparenleft}metis\ {\isacharparenleft}full{\isacharunderscore}types{\isacharparenright}\ eucl{\isacharunderscore}def{\isacharparenright}%
\endisatagproof
{\isafoldproof}%
\isadelimproof
\endisadelimproof
\begin{isamarkuptext}%
\begin{isbfig}{7em}
\begin{tikzpicture}[shorten >=1pt,node distance=2cm,on grid,auto] 
   \node[state] (i_1)   {$i_1$}; 
   \node[state] (i_2) [right=of i_1] {$i_2$}; 
   \node[state] (i_3) [right=of i_2] {$i_3$}; 
    \path[->] 
    (i_1) edge [loop above] node {} () 
          edge [bend left] node {} (i_2) 
    (i_2) edge [loop above] node {} () 
          edge [bend left] node {} (i_1) 
          edge [bend left] node {} (i_3) 
    (i_3) edge [loop above] node {} () 
          edge [bend left] node {} (i_2);
\end{tikzpicture}
\end{isbfig}%
\end{isamarkuptext}%
\isamarkuptrue%
\isamarkupsubsubsection{S5 $>$ B%
}
\isamarkuptrue%
\isacommand{lemma}\isamarkupfalse%
\ C{\isadigit{2}}{\isadigit{5}}{\isacharunderscore}A{\isacharcolon}\ {\isachardoublequoteopen}{\isasymforall}R{\isachardot}\ refl\ R\ {\isasymand}\ sym\ R\ {\isasymlongrightarrow}\ {\isacharparenleft}refl\ R\ {\isasymand}\ eucl\ R{\isacharparenright}{\isachardoublequoteclose}\isanewline
\ \isacommand{nitpick}\isamarkupfalse%
\isadelimproof
\ %
\endisadelimproof
\isatagproof
\isacommand{oops}\isamarkupfalse%
\endisatagproof
{\isafoldproof}%
\isadelimproof
\endisadelimproof
\isanewline
\isanewline
\isacommand{lemma}\isamarkupfalse%
\ C{\isadigit{2}}{\isadigit{5}}{\isacharunderscore}B{\isacharunderscore}ATP{\isacharunderscore}challenge{\isacharcolon}\ {\isachardoublequoteopen}{\isacharhash}\isactrlsup {\isadigit{3}}\ i{\isadigit{1}}\ i{\isadigit{2}}\ i{\isadigit{3}}\ {\isasymlongrightarrow}\ {\isasymnot}\ {\isacharparenleft}{\isasymforall}R{\isachardot}\ {\isacharparenleft}refl\ R\ {\isasymand}\ sym\ R{\isacharparenright}\ {\isasymlongrightarrow}\ {\isacharparenleft}refl\ R\ {\isasymand}\ eucl\ R{\isacharparenright}{\isacharparenright}{\isachardoublequoteclose}\isanewline
\isadelimproof
\ %
\endisadelimproof
\isatagproof
\isacommand{oops}\isamarkupfalse%
\ %
\isamarkupcmt{All ATPs time out%
}
\endisatagproof
{\isafoldproof}%
\isadelimproof
\isanewline
\endisadelimproof
\isacommand{theorem}\isamarkupfalse%
\ C{\isadigit{2}}{\isadigit{5}}{\isacharunderscore}C{\isacharcolon}\ {\isachardoublequoteopen}{\isacharhash}\isactrlsup {\isadigit{3}}\ i{\isadigit{1}}\ i{\isadigit{2}}\ i{\isadigit{3}}\ {\isasymand}\ r\ i{\isadigit{1}}\ i{\isadigit{1}}\ {\isasymand}\ r\ i{\isadigit{1}}\ i{\isadigit{2}}\ {\isasymand}\ {\isasymnot}\ r\ i{\isadigit{1}}\ i{\isadigit{3}}\ {\isasymand}\ r\ i{\isadigit{2}}\ i{\isadigit{1}}\ {\isasymand}\ r\ i{\isadigit{2}}\ i{\isadigit{2}}\ {\isasymand}\ r\ i{\isadigit{2}}\ i{\isadigit{3}}\ {\isasymand}\ {\isasymnot}\ r\ i{\isadigit{3}}\ i{\isadigit{1}}\ {\isasymand}\ r\ i{\isadigit{3}}\ i{\isadigit{2}}\ {\isasymand}\ r\ i{\isadigit{3}}\ i{\isadigit{3}}\ \ {\isasymlongrightarrow}\ {\isasymnot}\ {\isacharparenleft}{\isacharparenleft}refl\ r\ {\isasymand}\ sym\ r{\isacharparenright}\ {\isasymlongrightarrow}\ {\isacharparenleft}refl\ r\ {\isasymand}\ eucl\ r{\isacharparenright}{\isacharparenright}{\isachardoublequoteclose}\ \isanewline
\isadelimproof
\ %
\endisadelimproof
\isatagproof
\isacommand{by}\isamarkupfalse%
\ {\isacharparenleft}metis\ {\isacharparenleft}full{\isacharunderscore}types{\isacharparenright}\ eucl{\isacharunderscore}def\ refl{\isacharunderscore}def\ sym{\isacharunderscore}def{\isacharparenright}%
\endisatagproof
{\isafoldproof}%
\isadelimproof
\isanewline
\endisadelimproof
\isacommand{lemma}\isamarkupfalse%
\ C{\isadigit{2}}{\isadigit{5}}{\isacharunderscore}D{\isacharcolon}\ {\isachardoublequoteopen}{\isacharhash}\isactrlsup {\isadigit{2}}\ i{\isadigit{1}}\ i{\isadigit{2}}\ {\isasymlongrightarrow}\ {\isacharparenleft}{\isasymforall}R{\isachardot}\ {\isacharparenleft}refl\ R\ {\isasymand}\ sym\ R{\isacharparenright}\ {\isasymlongrightarrow}\ {\isacharparenleft}refl\ R\ {\isasymand}\ eucl\ R{\isacharparenright}{\isacharparenright}{\isachardoublequoteclose}\isanewline
\isadelimproof
\ %
\endisadelimproof
\isatagproof
\isacommand{by}\isamarkupfalse%
\ {\isacharparenleft}metis\ {\isacharparenleft}full{\isacharunderscore}types{\isacharparenright}\ refl{\isacharunderscore}def\ sym{\isacharunderscore}def\ eucl{\isacharunderscore}def{\isacharparenright}%
\endisatagproof
{\isafoldproof}%
\isadelimproof
\endisadelimproof
\isamarkupsection{Discussion and Future Work. \label{sec:eval}%
}
\isamarkuptrue%
\begin{isamarkuptext}%
The entire Isabelle document can be verified by Isabelle2014 in less than 60s on a semi-modern computer (2.4 GHz Core 2 Duo, 8 GB of memory).
When including all (commented) remote calls to the external ATPs in the calculation the verification time sums up to a few minutes,
which is still very reasonable. 

The improvements in comparison to the first-order based verification of the modal logic cube done
earlier by Rabe et al.~\cite{Rabe}, are: clarity and readability of the problem encodings, methodology,
reliability (our proofs are verifiable in Isabelle/HOL) and, most importantly, automation performance.
For the latter note that the experiments by Rabe et al.~\cite{Rabe} required several days of reasoning 
time in first-order theorem provers. Most importantly, however, their solution relied on an
enormous manual coding effort. However, we want to point again to the more general aims of their work.

Our solution instead requires a small amount of resources in comparison. In fact, as indicated before, 
the entire process (Steps A-D) is schematic, so that it should eventually be possible to fully automate our method.
For this it would be beneficial to have a flexible and accessible
conversion of the countermodels delivered by Nitpick back into Isabelle/HOL input syntax.
In fact, an automated conversion of Nitpick's countermodels into the corresponding \isa{C{\isacharasterisk}{\isacharunderscore}B} 
and \isa{C{\isacharasterisk}{\isacharunderscore}C} conjectures would eventually enable a truly automated exploration and verification of 
of the modal logic cube with no or minimal handcoding effort.
Similarly, for the interactive user a 
more intuitive presentation of Nitpick's countermodels would be welcome (perhaps similar to the illustrations we used 
in this paper).

Using the first-order provers E \cite{E}, SPASS \cite{SPASS}, Z3 \cite{Z3} and Vampire \cite{Vampire} proved unsuccessful for 
all \isa{C{\isacharasterisk}{\isacharunderscore}Isabelle{\isacharunderscore}challenge} problems (unless the right lemmas were given to them). Analyzing the reason for their weakness, as compared to the better performing higher-order automated theorem provers,
remains future work. In contrast, the SMT  solver CVC4 (via Sledgehammer) was quite successful 
and contributed five \isa{C{\isacharasterisk}{\isacharunderscore}Isabelle{\isacharunderscore}challenge} proofs.

Our work motivates further improvements regarding the integration of LEO-II and Satallax: While these systems
 are able to prove all \isa{{\isacharasterisk}{\isacharunderscore}Isabelle{\isacharunderscore}challenge} problems their proofs cannot yet be easily replayed or integrated 
in Isabelle/HOL. There have been recent improvements regarding the transformation of proofs from LEO-II and Satallax to 
Isabelle/HOL \cite{sultana14:_higher}, using which all the proofs produced by Satallax and LEO-II in
our work could be checked in Isabelle/HOL,\footnote{The proofs and the evaluation workflow can be downloaded from \url{http://christoph-benzmueller.de/papers/pxtp2015-eval.zip}}
but this process still requires some manual work to adapt the output from the ATPs.

Our work also motivates further improvements in higher-order automated theorem provers. For example, for these
systems it should be possible to also prove the remaining two \isa{{\isacharasterisk}{\isacharunderscore}ATP{\isacharunderscore}challenge} problems.
Moreover, they needed more than 10 seconds of CPU time in our experiments
for the \isa{{\isacharasterisk}{\isacharunderscore}Isabelle{\isacharunderscore}challenge} problems; it should be possible to prove these theorems much faster.%
\end{isamarkuptext}%
\isamarkuptrue%
\isamarkupsection{Conclusion \label{sec:conc}%
}
\isamarkuptrue%
\begin{isamarkuptext}%
We have fully verified the modal logic cube in Isabelle/HOL. Our solution is simple, elegant, easy to follow, effective 
and efficient. Proof exchange between systems played a crucial role in our experiments. In particular, we have exploited and combined Nitpick's 
countermodel-finding capabilities with subsequent calls to the higher-order theorem provers LEO-II and Satallax and the SMT solver
CVC4 via Isabelle's Sledgehammer tool. 
Our experiments also point to several improvement opportunities for Isabelle and the higher-order reasoners, in particular, 
regarding interaction and proof exchange.

Related experiments have been
carried out earlier in collaboration with Geoff Sutcliffe. Similar to
and improving on the
work reported in \cite{B12}, these unpublished experiments used the TPTP
THF infrastructure directly. However, in that work we did not achieve
a `trusted verification' in the sense of the work presented in this paper.
Another improvement in this article has been the use of schematic meta-level
working steps (Steps A-D) to systematically convert (counter)models found
by Nitpick into conjectures to be investigated. 

Future work will explore and evaluate similar logic relationships for other non-classical logics, for example, 
conditional logics. Any improvements in the mentioned systems, as motivated above, would be very beneficial
towards this planned work. Moreover, it would be useful to fully automate the schematic, meta-level working steps (Steps A-D) as
applied in our experiments. This would produce a system that would explore
logic relations truly automatically (for example, in conditional logics), analogous to what has been achieved here for the modal logic cube.%
\end{isamarkuptext}%
\isamarkuptrue%
\begin{isamarkuptext}%
\paragraph{Acknowledgements:} We thank Florian Rabe and the anonymous reviewers of this paper 
for their valuable feedback.%
\end{isamarkuptext}%
\isamarkuptrue%
\isadelimtheory
\endisadelimtheory
\isatagtheory
\endisatagtheory
{\isafoldtheory}%
\isadelimtheory
\endisadelimtheory
\end{isabellebody}%


\bibliographystyle{eptcs}
\bibliography{root}

\end{document}